\documentclass{jfm}
\usepackage{graphicx}
\usepackage{epstopdf, epsfig}
\usepackage{siunitx}
\usepackage{enumitem}
\usepackage{amsmath,latexsym}
\usepackage{subcaption}
\usepackage{comment}
\usepackage[retainorgcmds]{IEEEtrantools}

\DeclareSIUnit\Molar{M}

\shorttitle{Theory of hydrodynamic forces in electric double layers}
\shortauthor{C. Cramail; R. Lhermerout and E. Charlaix}

\title{Theory of hydrodynamic forces in electric double layers}

\author{C. Cramail, 
 R. Lhermerout
 \and E. Charlaix \corresp{\email{elisabeth.charlaix@univ-grenoble-alpes.fr}}}

\affiliation{Laboratoire Interdisciplinaire de Physique, Université Grenoble-Alpes, UMR 5588, 38402 Saint-Martin-d'Hères, France}

\begin{document}

\maketitle

\begin{abstract}
We present a theory of the hydrodynamic forces in the drainage flow of an electrolyte between a sphere and a plane with charged surfaces. Our theory considers a thin electrolyte film which is at all times in local equilibrium across its thickness. In addition to the usual lubrication force we explicit the electro-kinetic force due to the transport of the diffuse electrical charges as well as the diffusio-kinetic force due to the transport of the excess  ion concentration in the Electrostatic Double Layers (EDL’s). Our general formalism covers both the case of non-overlapping EDL’s and of overlapping EDL’s. We study more specifically the  mechanical impedance induced by an oscillatory motion of small amplitude of the surfaces. Among our main results, we show that the increase in damping due to the electro-kinetic effect is not monotonic with the surface charge, and we predict a diffusio-kinetic stiffness with a long range decay in $1/D^4$  susceptible to overcome the  stiffness of the equilibrium Derjaguin-Landau-Verwey-Overbeek force. The comparison of our results with experiments could allow one to confront the theories of electrolyte transport in the Electrostatic Double Layers without ajustable parameters. 
\end{abstract}


\section{Introduction}
The electrostatic and hydrodynamic properties of the solid-liquid interface are at the core of a wide range of biological processes and industrial applications, such as those involving the stability of colloidal suspensions or membrane filtration. This ubiquity justifies the abundance of research conducted on this topic for over a century, at the crossroads of physics, chemistry, and biology. In the context of the necessary decarbonization of our energy sources and growing tensions over freshwater access, a better understanding of equilibrium and transport at the solid-liquid interface has once again become crucial \citep{Bocquet2010, Siria2017, Bocquet2020, Bonn2021}. \\
In the early 1850s, Helmholtz was the first to describe the charge separation occurring at the interface between a dielectric solid and an ionic solution. He introduced the term electric double layer (EDL) to refer to the electrical capacitance formed between the charged surface and the counterions in the interfacial layer of the solution. Later, in the early 1910s, Gouy and Chapman independently derived the distributions of electrostatic potential and ionic concentrations at the interface by solving the so-called Poisson-Boltzmann equation, which predicts their exponential decay away from the interface. The DLVO equilibrium force – named after Derjaguin, Landau, Verwey, and Overbeek – was proposed 30 years later based on Gouy-Chapman (GC) theory to describe the stability of colloidal suspensions, with significant industrial implications. Although widely used, the DLVO theory – and, by extension, GC theory – was only experimentally tested in the late 1970s using the Surface Force Apparatus (SFA), which had been invented a decade earlier by Tabor, Winterton, and Israelachvili \citep{Israelachvili2011, Tabor1969, Israelachvili1978, Horn1989}. Measurements showed that GC theory accurately describes reality beyond a few nanometers from the dielectric surface when the solution is sufficiently dilute, typically at concentrations below \SI{1}{\Molar}. \\
In the molecular layer adjacent to the surface, which deviates from Gouy-Chapman behavior and is traditionally referred to as the Stern layer, capacitance measurements instead suggest a linear decay of the electrostatic potential. Unlike the so-called diffuse layer, which is well described by GC theory, the Stern layer remains largely enigmatic \citep{Bourg2017, You2018, Atkin2024}. Today, the term EDL is commonly used to describe this division of the interfacial solution into two layers based on the validity of GC theory, but here, we will use it in Helmholtz’s original sense, excluding the Stern layer from our discussion. \\
So far, lateral transport at the interface has mainly been described using classical transport equations (with the mean-field approximation for electrostatic interactions) – the Stokes equation for the liquid and the Nernst-Planck equation (with advection) for the ions – within the electric double layer. Smoluchowski, a pioneer in the quantitative description of electrokinetic phenomena, originally derived an expression for the electroosmotic flow rate and streaming current by applying the Stokes equation within the electric double layer, even though he did not know the equilibrium distributions at the time, in 1903 \citep{Smo1924}. Unlike Smoluchowski, expressions for diffusio-electrokinetic interfacial transport (coupled transport of volume, charge, and solute) have been used for over a century, particularly in colloidal science, assuming equilibrium in the direction normal to the interface, governed by GC theory, with a surface charge identical to that of the overall equilibrium \citep{Gross1968, Prieve1984}. For instance, Smoluchowski’s expression for electrophoretic velocity is used to measure a zeta potential, equal to the surface electrostatic potential in the absence of slip, from which a surface charge is deduced using Grahame’s equilibrium relation (1953), which follows from GC theory \citep{Lyklema1994, Bonthuis2012, Siria2013, Hartkamp2018}. However, the surface charge thus deduced is never directly compared to a direct measurement of the GC surface charge at global equilibrium. Furthermore, the no-slip boundary condition is potentially used inappropriately, as the hydrodynamic boundary condition is rarely measured \citep{Bocquet2007}. \\
Yet, the surface charge deduced from electrophoresis measurements using the aforementioned classical theory can differ by several orders of magnitude from the intrinsic surface charge measured by titration. Surface charge measurements based on conductivity, which rely on the same classical theoretical framework, sometimes yield values even more divergent from the other two \citep{Lyklema1994, Bonthuis2012, Siria2013, Hartkamp2018}. In 2012, Netz and collaborators proposed an interpretation for the multiplicity of experimental surface charges, inspired by molecular dynamics simulations. Using viscosity and dielectric constant profiles defined by simple Heaviside functions (step-like profiles) from the surface, the authors reproduced the relative experimental behaviors of the three aforementioned surface charges. However, this avenue requires further investigation, particularly through additional experimental tests. \\
Despite the experimental advancements since Smoluchowski, which have brought increasingly greater sensitivity to surface effects, the classical description of diffusio-electrokinetic phenomena still lacks direct experimental validations. We believe that dynamic SFA and colloidal probe AFM (CP-AFM) measurements, combined with equilibrium measurements, could address this gap. The SFA is indeed the instrument of choice for measuring GC surface charge based on DLVO theory. Moreover, its dynamic version, along with its counterpart, the CP-AFM, has demonstrated its ability to measure viscous forces and hydrodynamic boundary conditions \citep{Cottin-B2005, Maali2008, Garcia2016, Lizee2024}. However, a model describing diffusio-electrokinetic interfacial transports during drainage of a thin film of an ionic solution in SFA or CP-AFM is still lacking. \\
To be more precise, several authors have investigated the hydrodynamic forces exerted upon drainage of a thin film of ionic solution in sphere-plan geometry. Their various studies predict that these forces have a purely dissipative nature, originating from electro-viscous effects.
In 1990, Bike and Prieve calculated, at large sphere-plane separations, the first correction term to the Reynolds force, which results from the pressure-driven flow within the liquid film \citep{Prieve1990}. More recently, Würger and collaborators analytically derived all subsequent correction terms \citep{Rodriguez2022}. However, their model does not provide an analytical expression for the force outside this asymptotic regime. Moreover, these authors entirely neglect concentration gradients and ion transport by diffusion – an assumption that requires further validation.
Finally, in 2020, Mugele and collaborators developed a model to explain their observations of electro-viscous forces in dynamic AFM. In their model, ion transport by convection is considered negligible compared to transport by diffusion and migration – an assumption valid in their case, as they worked with an AFM tip of approximately \SI{1}{\micro \meter} in radius \citep{Mugele2020, Liu2018}. However, this model is not suitable for describing the transport mechanisms involved in dynamic SFA and certain CP-AFM with larger probes. \\
In this paper, we propose a model based on the aforementioned classical framework for the normal hydrodynamic forces measured in SFA or CP-AFM, in the linear response regime, during drainage of a thin film of an ionic solution. The film is considered to be at equilibrium along its thickness. Coupled transports of volume, charge, and solute are all taken into account. Our model is built on a single surface charge: the equilibrium surface charge as described by the Gouy-Chapman electric double layer theory. However, considering a surface charge that varies with the thickness of the film is not an obstacle. The no-slip hydrodynamic boundary condition at the walls is also adopted.\\
The paper is organized as follows. Sec. 2 presents the equations governing the drainage of a thin film of ionic solution and derives the total force within the Derjaguin approximation. These equations are then simplified in Sec. 3 by assuming local equilibrium across the film thickness. In Sec. 4, we calculate the response force to a surface harmonic oscillation  in the linear response regime, and we identify the different physical mechanisms contributing to the mechanical impedance. In Sec. 5, we analyse the contribution of the electro-kinetic effects, while Sec. 6 focuses on the forces generated by diffusio-kinetic mechanisms. Finally, Sec. 7 establishes our conclusions. 


\section{Electro-hydrodynamic forces in the sphere-plane geometry: the Derjaguin approximation}\label{sec:2}
\subsection{General framework of the model}
\label{subsubsec:principle_intro}
Our model describes applications such as dynamic SFA or colloidal probe AFM.
We consider two  solid spheres of same radius $2R$, physically identical to each other,  located at a small distance $D\ll R$ the one from the other. In the frame of the Derjaguin approximation\cite{} and related approximations developped below, this geometry is equivalent to a sphere of radius $R$ and a plane.  We assume a full symmetry around the axis $Oy$ defined by the sphere centers with the origin $O$ located at mid-distance of the two spheres, as well as a plane symmetry across $y=0$, and we use the cylindrical coordinates $(O,r,y)$. The gap between the surfaces at a distance $r$ of the $Oy$ axis is noted $z$. The system defined in this manner is depicted in Fig. \ref{schema_geometrie}.

The solids are immersed in a solution of monovalent electrolyte of number density $n_o$ (number of cations and anions per unit volume) and their surface are assumed to bare a uniform surface charge $\sigma$.  The spheres are oscillated harmonically in the normal direction around their nominal position, so that the gap $D(t)$ is: 
$$D(t) = D + h_0 \cos (\omega t) \qquad h_0 \ll D \ll R$$
The quantity of interest is the hydrodynamic force $F$ normal to the surfaces, and more specifically the linear force response with respect to the oscillation amplitude $h_0$: 
$$Z(D,\omega)=-\frac{F_{\omega}}{h_0} \qquad F(D,t)=F_{eq}(D)+\Re[F_{\omega}e^{i\omega t}]
+o(h_0^2)$$
where  the suffix $eq$ refers to the equilibrium situation ($h_0=0$).  We choose to use a negative sign to define $Z$ in order to get positive real and imaginary part for usual stiffness and dissipative components. We will consider here only the contributions to the force related to hydrodynamics and electrokinetics, at the exclusion of  Van der Waals forces and solvation forces, and we will neglect the possible elastic deformation of the solid surfaces. In these conditions  the complex amplitude $Z(D,\omega)$ defines the electro-hydrodynamic impedance of the system. 
\setlength{\unitlength}{1cm}
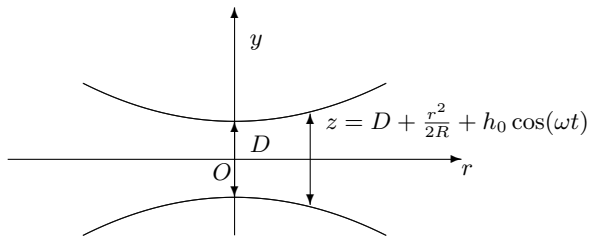
\begin{figure}
\begin{center}
\begin{picture}(6,5)(-3,-2)
\put(-3,0) {\vector (1,0){6}}
\put(3,-0.2) {$r$}
\put(0,-1) {\vector (0,1){3}}
\put(0.2,1.5) {$y$}
\qbezier(-2,1)(-1,0.5)(0,0.5)
\qbezier(0,0.5)(1,0.5)(2,1)
\qbezier(-2,-1)(-1,-0.5)(0,-0.5)
\qbezier(0,-0.5)(1,-0.5)(2,-1)
\put(0,0) {\vector (0,1){0.5}}
\put(0,0) {\vector (0,-1){0.5}}
\put(0.2,0.1) {$D$}
\put(-0.29,-0.27) {$O$}
\put(1,0) {\vector (0,1){0.6}}
\put(1,0) {\vector (0,-1){0.6}}
\put(1.2,0.4) {$z=D+\frac{r^2}{2R} + h_0 \cos(\omega t)$}
\end{picture}
 \caption{Schematics of the system geometry: an electrolytic solution is confined between two identical spheres of total curvature $1/R$ and whose separation distance is oscillating.}
\label{schema_geometrie}
\end{center}
\end{figure}

Our model uses the standard formulation of the electrokinetic theory in the regime of slowly varying fields \cite{}, that is neglecting the magnetic fields and wave propagation. The transport of ions is governed by the Poisson-Nernst-Planck equation,
and we will note $n^+$,  $n^-$  the number density of cations and anions, $D^+$, $D^-$ there diffusion coefficient, and $\mu^+=k_BT \ln n^+/n_o+eV$, $\mu^-=k_BT \ln n^-/n_o-eV$ their electro-chemical potential, where $T$ is the temperature, $k_\mathrm{B}$ the Boltzmann constant, and $e$ the elementary charge. We will also use the non-dimensional electrical potential $\psi=eV/k_BT$. 

In SFA and AFM applications, the Reynolds number  is essentially very low. An upper limit is $Re < h_o\omega R/\eta$ with $\eta$ the liquid viscosity. With an amplitude $h_o$ limited to 10nm, a frequency of the kHz, a millimetic radius of curvature and the viscosity of water, the Reynolds hardly reaches $10^{-4}$. The flow  will thus be described by the incompressible Stokes equation, which we write using the  hydrodynamic stress tensor: 
\begin{equation}
\mathsfbi{S} ^{\mathcal H} = -P\bar{\bar I}  + \frac{\eta }{2} (\bar{\bar \nabla }\vec v+^t \bar{\bar \nabla}\vec v)  \qquad \vert y \vert < z/2
\end{equation}
where $ \vec v$ is the flow velocity, $\bar{\bar I}$ the identity matrix, and $P$ the pressure in excess to the atmospheric pressure.  
We will also use 
the electrostatic Maxwell stress tensor:
\begin{equation}
\mathsfbi{S}^{\mathcal M} = \varepsilon_i\left (  - \frac{\Vert \vec \nabla V \Vert ^2}{2}\bar{\bar I}+\vec \nabla V \otimes \vec \nabla V \right ) \qquad
\begin{cases}
    \varepsilon_i=\varepsilon_l \ \ {\rm if} \ \vert y \vert < z/2 \\
    \varepsilon_i=\varepsilon_s \ \ {\rm if} \ \vert y \vert > z/2
\end{cases}
\end{equation}
where 
$\varepsilon_l$, $\varepsilon_s$ are the dielectric permittivities respectively of the liquid and of the solids.
With these notations the conservation laws and Poisson-Nernst-Planck-Stokes equations write:
\begin{IEEEeqnarray}{rCl}
\nabla^2 V = - \frac{e(n^+-n^-)}{\varepsilon_l} \quad \vert y \vert < \frac{z}{2} \qquad 
    \nabla^2 V =0 \quad \vert y \vert > \frac{z}{2} 
    \IEEEyesnumber \IEEEyessubnumber
 \label{eq_Poisson}\\
   \frac{\partial n^{\pm}}{\partial t} ={\rm div} j^{\pm} \qquad {\rm div} \vec v = 0 
    \IEEEyessubnumber
   \label{eq_localconservation}\\
 j^{\pm}=-\frac{D^{\pm}}{k_BT} n^{\pm} \vec \nabla \mu^{\pm} + n^{\pm} \vec v  
  \IEEEyessubnumber
 \label{eq_Nernst_Planck}\\
 {\rm div }(\mathsfbi{S} ^{\mathcal M}+\mathsfbi{S} ^{\mathcal H}) = \vec 0 = -\vec \nabla P + \eta \nabla^2 \vec v - e (n^+-n^-) \vec \nabla V
  \IEEEyessubnumber
 \label{eq_Stokes}
\end{IEEEeqnarray}

The boundary conditions at the symmetry plane $y=0$ also noted with the subscript m, correspond to invariance by the transformation $y \rightarrow -y$:
\begin{equation}
y=0  \ :\quad 
0=v_y=\frac{\partial V}{\partial y}=\frac{\partial \mu^{\pm}}{\partial y} = \frac{\partial v_r}{\partial y} =0 
\label{eq_bcsym}
\end{equation}
At the liquid solid interface $y=z/2$ also noted with the subscript w, we will assume:
i) a no-slip boundary condition, 
ii) no ion source, 
iii) the continuity of the electrical potential and the Maxwell-Gauss equation for the normal component of the electrical displacement. These boundary conditions at the solid interface write: 
\begin{IEEEeqnarray}{rCl}
\vec v_{\rm{s}}=-h_0 \omega \sin \omega t \ \vec e_y 
 \IEEEyesnumber \IEEEyessubnumber\\
\vec \nabla \mu^{\pm}\vert_{{\rm w}} \cdot \vec n  = 0
\IEEEyessubnumber\label{eq_bcm} \\
V_{{\rm w}^-}=V_{{\rm m}^+}=V_{\rm w}
\IEEEyessubnumber \label{eq_psi} \\
- \varepsilon_s \vec \nabla V \vert_{{\rm w}^+} \cdot \vec n + \varepsilon_l \vec \nabla V \vert_{{\rm w}^-}\cdot\vec n= \sigma
\IEEEyessubnumber \label{eq_bcE}
\end{IEEEeqnarray}
In the following we will ignore charge regulation effects and consider that $\sigma$ is uniform and non variable. Without loss of generality we will set sgn($\sigma$)=1. 

Finally the total stress is continuous through the interface. The force $F=\vec F \cdot \vec e_y$ acting on the upper sphere is obtained formally from the force balance:
\begin{equation}
\vec F \cdot \vec e_y -\int_0^{\infty} \vec e_y \cdot(\mathsfbi{S} ^{\mathcal M}_{{\rm w}^+} \ \vec n)\  2\pi rdr = - \int_0^{\infty} \vec e_y \cdot \left [(\mathsfbi{S} ^{\mathcal H}+\mathsfbi{S} ^{\mathcal M} )\vert_{{\rm w}^-} \ \vec n\right  ]  \ 2\pi rdr
\label{eq_formalforce}
    \end{equation}
Our models calculates the linear force response within two further approximations: \\
$\bullet$ the Derjaguin approximation and its consequences discussed in next paragraph\\
$ \bullet$ the hypothesis of local thermodynamic equilibrium across the thickness $z$ of the liquid film, developped in section \ref{sec:3}

\subsection{The Derjaguin approximation}
The Derjaguin approximation assumes that the force acting on  surface elements at a distance $z$ of each other becomes negligible when $z\gtrsim \lambda \ll R$. The characteristic lengths setting the magnitude of $\lambda$, are the Debye's length $\ell_D$ which sets  the range of electrostatic interactions, and the smallest distance $D$ between the surfaces, which sets the range of the hydrodynamic interactions. 
In these conditions 
the interacting surfaces are always almost parallel to each other, so that $\vec n \approx \vec e_y$, and $\vec n$ has to be replaced by $\vec e_y$ in the boundary conditions (\ref{eq_bcm}, \ref{eq_bcE},\ref{eq_formalforce}). Also according to the Derjaguin approximation, the gap between the surfaces reduces to the parabolic approximation:
\begin{equation}
z(r, t) = z(r) + h_0 \cos(\omega t) \qquad z(r)=D + \frac{r^2}{2R}
\label{approx_parabol}
\end{equation} 
Hence the Derjaguin approximation describes as well a sphere of radius $R$ and a plane.

Because of the hypothesis $\lambda \ll R$,  the variation of the physical quantities in the liquid film along the $r$ axis occurs on a radial distance of order $\sqrt{R\lambda}$  much larger than the distance of variation along the $Oy$ axis, which is of order of some   $\lambda$. Thus for any quantity $a$ which varies significantly across the thickness the liquid film we have:
\begin{equation} \left \vert \frac {\partial a}{\partial r } \right \vert \ll \left \vert \frac {\partial a}{\partial y } \right \vert \qquad \vert y \vert \le \frac{z}{2}
\label{eq_27}
\end{equation}

A first important consequence of the Derjaguin approximation is that  the Poisson's equation in the liquid (\ref{eq_Poisson}, $\vert y \vert < z/2$) and its boundary condition (\ref{eq_bcE}) become:
\begin{equation}
\frac{\partial^2 V}{\partial y^2} = -\frac{e (n^+-n^-) }{\varepsilon_{\ell}} \qquad \frac{\partial V}{\partial y}\left [ \left (\frac{z}{2} \right ) ^-\right ] = \frac{\sigma}{\varepsilon_{\ell}}
\label{eq_Poissonaxial}
\end{equation}
It is indeed shown in the Appendix that the normal field in the dielectric  is of the same order of magnitude as the radial field $(\partial V/\partial r) [(z/2)^-]$ which is neglected. By integrating eq. (\ref{eq_Poissonaxial}) along the $y$-direction one sees that the Derjaguin approximation implies the local electroneutrality of the electrolyte film together with the surfaces:
\begin{equation}
 \quad \int_0^{z(t)/2}e (n^+-n^-)(r,y,t)dy=-\sigma
\label{eq_electroneutrality}
\end{equation}

As the surfaces are almost parallel to each other in the interaction region,  $\vec n \approx \vec e_y$. Together with the inequality (\ref{eq_27}), the normal component of the Maxwell stress tensor at the interface reduces to:
$$ \mathsfbi{S} ^{\mathcal M}_{{\rm w}} \ \vec n \approx S^{\mathcal M}_{yy} \vec e_y 
\approx \frac{\varepsilon_i}{2}\left (\frac{\partial V }{\partial y } \right)^2 \vec e_y$$
It is thus negligible in the dielectric. The expression of the force (\ref{eq_formalforce}) becomes:
\begin{align}
F  & = - \int_0^{\infty}2\pi r (S ^{\mathcal M}_{yy}+S ^{\mathcal H}_{yy} ) (r,\frac{z^-}{2}) dr \nonumber \\
&=  - \int_0^{\infty}2\pi r \left (-P+2\eta \frac{\partial v_y}{\partial y} +\frac{\varepsilon_{\ell}}{2}\left (\frac{\partial V }{\partial y } \right)^2 \right )(r,\frac{z}{2}^-) dr
\label{eq_215}
\end{align}
We show in appendix that within the Derjaguin approximation the force reduces to:
\begin{equation}
    F=\int_0^\infty 2\pi r P(r,0)dr 
    \label{eq_force}
\end{equation}


\subsection{Equilibrium state}
In the following we will linearize the dynamic force to the first order in $h_0$ around the equilibrium state. The latter has been described by \cite{} and leads to the well-known DLVO force \cite{}. At equilibrium the electro-chemical potential of the ions are uniform: $\mu_{eq}^{\pm}=0$. The ion densities obey the Boltzmann law $n_{eq}^{\pm}=n_o e^{\mp \psi_{eq}}$ where the reduced potential $\psi_{\rm eq}=eV_{\rm eq}/k_BT$ is governed by the Poisson-Boltzmann equation:
\begin{equation}
    \frac{\partial ^2\psi_{eq}}{\partial y^2}=\frac{\sinh \psi_{eq}}{\ell_D^2} \qquad \ell_\mathrm{D}^2 = \frac{\varepsilon_l k_\mathrm{B} T}{2e^2 n_0}=\frac{1}{8\pi \ell_B n_o}
    \label{eq_Debyelength}
\end{equation}
and $\ell_D$ is the Debye's length. 
The analytical solution for $\psi_{eq}$ is \citep{Behrens2001}:
\begin{equation}
\psi_{eq}(r,y)=\psi_{\rm eq,m}(r)-\ln \left [{\rm cd}\left (y/2 \ell_D \sqrt{k} \right ) \right ]^2 \qquad k=e^{-\psi_{eq,m}}
\label{eq_jacobi-potential}
\end{equation}
for a positive surface charge. The subscript m refers to the middle position $y=0$ and
cd($u\vert k^2$) is the Jacobi elliptical function of argument $u$
and parameter $k^2$. The parameter $k^2(z)$ is determined from the boundary condition (\ref{eq_Poissonaxial}) by:
\begin{eqnarray}
2 \cosh  \psi_{eq,s}=k{\rm cd}^2(z'/2)+\frac{1}{k{\rm cd}^2(z'/2)}=k+\frac{1}{k}+\alpha^2
\\ z'=\frac{z}{2\ell_D\sqrt{k}} \qquad \alpha = \frac{2l_D}{l_G} \nonumber
\label{eq_keq}
\end{eqnarray}
We note that $k$ depends only on $(z/\ell_D, \ell_G/\ell_D)$, and the function $\psi_{eq}(y)$ depends on  $(y/\ell_D,z(r)/\ell_D,\ell_G/\ell_D)$. Finally the equilibrium pressure is obtained by projecting on the $r$ axis the hydrostatic equation (\ref{eq_Stokes}) with $\vec v= \vec 0$:
$$\frac{\partial P_{eq}}{\partial r} = 2 k_BT n_o \sinh \psi_{eq} \frac {\partial \psi_{eq}}{\partial r}
$$
With the value $P_{eq}=\psi_{eq}=0$ in the reservoir, one finds that
 $P_{eq}=2n_ok_BT (\cosh \psi_{eq}-1)$,
and equation (\ref{eq_force}) retrieves the equilibrium DLVO force:
$$
  F_{eq}= 2n_ok_BT \int_0^\infty 2\pi r (\cosh \psi_{eq,m}-1) dr
$$  
where $\psi_{eq,m}=\psi_{eq}(y=0)$ is the reduced potential in the symmetry plane. Using the parabolic approximation $z=D+r^2/2R$ the equilibrium force is written in the usual form
\begin{equation}
  F_{eq}= 4\pi Rn_ok_BT \int_D^\infty (\cosh \psi_{eq,m}-1) dz
  \label{eq_eqforce}
\end{equation}
\section{Hypothesis of local equilibrium across the film thickness}
\label{sec:3}
\subsection{Conditions}
Due to the separation of scales, we assume that equilibrium is instantaneously reached along the confinement axis.  This assumption is based on the comparison of the working period $2 \pi /\omega$ with the relevant ionic diffusion time $\tau_y$ normal to the surfaces. A lower limit of $\tau_y$ is estimated for a Debye's length $\ell_D \sim 100$ nm, as: 
\begin{align}
    \tau_y & \sim \frac{\lambda_\mathrm{D}^2}{D_\mathrm{diff}} \sim \frac{(10^{-7})^2}{10^{-9}} = 10^{-5} \, \mathrm{s} 
\end{align}
This assumption therefore holds true in dynamic SFA with a typical working frequency of $\SI{5}{\hertz}$ to $\SI{500}{\hertz}$. However in colloidal probe AFM,  the working frequency should not exceed $\SI{1}{\kilo\hertz}$.

\subsection{Thermodynamic potentials}
Under this local equilibrium assumption, the electrochemical potential $\mu^+$ and $u^-$ of the ions are uniform across the thickness of the electrolyte film. 
We introduce the sum and difference potentials as well as the equivalent concentration $n(r,t)$ as:
\begin{IEEEeqnarray}{rCl}
  \mu_s=\frac{\mu^++\mu^-}{2}=k_BT\ln\frac{n(r,t)}{c_o}
     \qquad  W=\frac{\mu^+-\mu^-}{2e}  \qquad
\IEEEyesnumber 
    \label{eq_32}     \IEEEyessubnumber \\
        \chi(r,y,t)  =\frac{e}{k_BT }(V-W)  \quad \IEEEyessubnumber
\end{IEEEeqnarray}
The equivalent ion density $n(r,t)$ is the ion density of a virtual reservoir at 0-potential that would be in equilibrium with the film at radial distance $r$ and time $t$. The sum and difference concentration of ions are defined as:
\begin{equation}
 n_s=n^++n^-=2n(r,t)\cosh \chi \qquad n_e=n^+-n^-=-2n(r,t)\sinh \chi   \label{eq_csce}
\end{equation}
The potentials $\mu_{\rm s}$ and $W$  are uniform over the thickness. Therefore the reduced potential $\chi$ satisfies the Poisson equation (\ref{eq_Poissonaxial}) with the expression (\ref{eq_csce}) for the charge density:
\begin{equation}
    \frac{\partial ^2\psi}{\partial y^2} = \frac{\partial ^2\chi}{\partial y^2} = \frac{n(r,t) \sinh \chi}{\ell_D^2}
\qquad \frac{\partial \chi}{\partial y}\vert_{z(r,t)/2}=\frac{\partial \psi}{\partial y}\vert_{z(r,t)/2}=-\frac{2}{\ell_G}
\label{eq_34}
\end{equation}
This set of equation and boundary condition is equivalent to the one defining the equilibrium electrical potential $\psi_{eq}$, but with a local Debye's length $\lambda_D(r,t)$ and a boundary condition applied at the moving wall $z(t)$. The solution is thus:
\begin{equation}
    \chi(r,y,t)=\psi_{eq}(y,z(r,t),\lambda_D) \qquad \lambda_D(r,t)=\frac{\ell_D}{\sqrt{n(r,t)}}=\ell_D e^{-\mu_s(r,t)/2k_BT}
    \label{eq_chi}
\end{equation}
This means that the dynamic concentration profiles (\ref{eq_csce}) reach instantaneously the equilibrium value that would prevail if the reservoir concentration was $n(r,t)=n_oe^{\mu_s/kBT}$ and the distance between the surface $z(t)$. 
 
The transport equations are written in terms of the sum and difference variables by introducing the average diffusion coefficient:
\begin{equation}
    D_s=\frac{D^++D^-}{2} \qquad \delta = \frac{D^+-D^-}{D^++D^-} \qquad D^{\pm}=D_s(1\pm \delta) 
\end{equation}
Using equations (\ref{eq_32}) and (\ref{eq_csce}) we get the identity:
\begin{equation}
e(n^+-n^-) \vec \nabla V=en_{\rm e} \vec \nabla W +n_s\vec \nabla \mu_s-k_BT \vec \nabla n_s 
\label{eq_38}
\end{equation}
and write eq (\ref{eq_localconservation}, \ref{eq_Nernst_Planck}) and (\ref{eq_Stokes}) as:   
\begin{IEEEeqnarray}{rCl}
\frac{\partial n_s}{\partial t }= - {\rm div} \vec j_s \qquad 
\frac{\partial n_e}{\partial t }= - {\rm div} \vec j_e \qquad
{\rm div} \vec v = 0 
    \IEEEyesnumber  \label{eq_37}
    \IEEEyessubnumber \label{eq_37a} \\
 \vec j_s=n_s \vec v - \frac{D_s}{k_BT}\left \{ (n_s+\delta n_e)\frac{d\mu_s}{dr} + e(n_e+\delta n_s) \frac{dW}{dr} \right \}
 \IEEEyessubnumber  \label{eq_37b}\\
\vec j_e=n_e \vec v - \frac{D_s}{k_BT}\left \{(n_e+\delta n_s) \frac{d\mu_s}{dr} + e(n_s+\delta n_e) \frac{dW}{dr} \right \}
\IEEEyessubnumber  \label{eq_37c}\\
\eta\frac{\partial ^2 v_r}{\partial y^2}  = \frac{d \Pi}{d r} + en_e \frac{d W}{d r}+(n_s-2n) \frac{d \mu_s}{d r}
\IEEEyessubnumber  \label{eq_37d}\\
\Pi=  P-(n_s-2n)k_BT
\IEEEyessubnumber  \label{eq_37e}
\end{IEEEeqnarray}
In the Stokes equation (\ref{eq_Stokes}) the radial components $\Delta_r \vec v$  has been neglected  according to the Derjaguin approximation, to obtain (\ref{eq_37d}). We show in the Appendix that the osmotically compensated  pressure  $\Pi$  defined in (\ref{eq_37e}) is uniform across the film thickness.  

The three terms of the r.h.s. of the Stokes equation (\ref{eq_37d}) outline the three contributions to the flow. The first term $d\Pi/dr$, uniform in the section, induces a Poiseuille-like velocity profile: this is the pressure driven flow. The second term is the effective electrical driving force, as -$dW/dr$ is the radial electrical field which develops across the liquid film. This term vanishes outside of the EDLs as the volume charge $en_e$, it is the source of an electro-osmotic flow.  The third term also eventually vanishes outside of the EDLs, it induces a diffusio-osmotic flow driven by the concentration gradient $d\mu_s/dr = (k_BT/n) dn/dr$ (\cite{Prieve1984}).

 \subsection{Conservation laws integrated over the thickness}
The above equations are integrated across the thickness by defining: 
 \begin{eqnarray}
 N_s(r,t)=\int_0^{z(t)/2}(n_s-2n_o) dy \qquad 
 N_e(r,t)=\int_0^{z(t)/2} n_e dy = -\frac{\sigma}{e} 
 \label{eq_NsNe}\\
 J_{\mathrm{s},\mathrm{e}}= \int_0^{z(t)/2} \vec j_{s,e}\cdot \vec e_r dy  
 \qquad J_\mathrm{v}=\int_0^{z(t)/2}v_r(y) dy 
 \end{eqnarray}
 The constant value of $N_e$ reflects the global electro-neutrality  (\ref{eq_electroneutrality}). 

In view of equations (\ref{eq_37}) it is clear that the  fluxes  $(J_\mathrm{v},J_{\mathrm{s}},J_{\mathrm{e}})$  are linear functions of the gradients of the thermodynamic potential $ (\Pi,\mu_s,W)$. 
 We show in the appendix that these fluxes are related to the  above gradients by the 3x3 symmetric matrix $\mathsfbi{T}$ as: 
\begin{IEEEeqnarray}{rCl}
\begin{pmatrix} J_{\rm v}\\ J_{\rm s}-2n(r,t)J_{\rm v} \\eJ_{\rm e}\end{pmatrix} = -\mathsfbi{T} \frac{\partial }{\partial r} \begin{pmatrix} \Pi  \\\mu_s \\ W  \end{pmatrix}
\qquad \mathsfbi{T} =\frac{\mathsfbi{K} }{\eta}
+\frac{D_s\mathsfbi{L} }{k_BT}
\IEEEyesnumber \label{eq_matrix} \IEEEyessubnumber\label{eq_matrixflux}\\
K_{ij}= \int_0^{z(t)/2} \Gamma_i\Gamma_j dy  \qquad (i,j)\in (\mathrm{v},\mathrm{s},\mathrm{e})
 \IEEEyessubnumber\label{eq_Kij}\\
\Gamma_\mathrm{s} (y)= \int_0^y \left ( n_s -2n\right)\mathrm{d}y \qquad  \Gamma_\mathrm{e} (y)= \int_0^y en_{\rm{e}} \mathrm{d}y \qquad \Gamma_\mathrm{v} (y)  = y 
\IEEEyessubnumber\\
L_{ \mathrm{ve}}=L_{ \mathrm{ev}}=L_{ \mathrm{vs}}=L_{ \mathrm{sv}}=0  \IEEEyessubnumber
\\
L_{\mathrm{s}\mathrm{s}}=\frac{L_{\rm ee}}{e^2}=
N_s+\frac{z}{2}+ N_e \delta\qquad L_{\mathrm{s}\mathrm{e}}=L_{\mathrm{e}\mathrm{s}}=e\left [N_e+ (N_s+\frac{z}{2})\delta \right ]
\IEEEyessubnumber\label{Gamma_def}
\end{IEEEeqnarray}
The two components $\mathsfbi{K}$ and $\mathsfbi{L}$ of the transport matrix $\mathsfbi{T}$ reflect the convective and the diffusive contributions in the flux expressions (\ref{eq_37b}) and (\ref{eq_37c}).

The three conservation laws (\ref{eq_37a}) integrated  over the thickness write as:
 \begin{IEEEeqnarray}{rCl}
  \frac{\partial (z/2)}{\partial t}=\frac{\dot D }{2}= -  \frac{1}{r} \frac{\partial r J_\mathrm{v}}{\partial r} \qquad \qquad
 \IEEEyesnumber \label{eq_312} \IEEEyessubnumber \label{eq_312a}\\
 \frac{\partial N_{\rm e}}{\partial t}=\int_0^{z(t)/2}\left (\frac{\partial c_{e}}{\partial t } +\frac{\partial j_{e,y}}{\partial y }\right )dy = - \frac{1}{r} \frac{\partial r J_{e}}{\partial r} \qquad \qquad
 \IEEEyessubnumber\\ 
 \frac{\partial N_{s}}{\partial t}=\int_0^{z(t)/2}\left (\frac{\partial c_{s}}{\partial t } +\frac{\partial j_{s,y}}{\partial y }\right )dy -\frac{\partial z}{2\partial t} = - \frac{1}{r} \frac{\partial r (J_{s}-2nJ_{\rm v})}{\partial r} -\frac{2}{r} \frac{\partial r (n-n_o)J_{\rm v}}{\partial r}  
  \qquad \qquad
 \IEEEyessubnumber \label{eq_312c}
 \end{IEEEeqnarray}
 
 These equations are further reduced by noting that $\partial z/\partial t=-h_0\omega \sin \omega t$ does not depend on $r$, therefore the first equation can be integrated once, leading to $J_{\rm v}=-(r/4) (\partial z/\partial t)$.  Furthermore in the absence of charge-regulation phenomena $N_e$ is constant (see eq. \ref{eq_electroneutrality}), and the second  equation integrates in $J_e=0$ as $J_e$ does not diverge at $r=0$. Using the relations (\ref{eq_matrixflux}) and eliminating $dW/dr$,  equations (\ref{eq_312a}) and (\ref{eq_312c}) become a set of two second-order non-linear differential equations:
\begin{IEEEeqnarray}{rCl}
\left (T_{\rm vv}-\frac{T_{\rm ev}^2}{T_{\rm ee}} \right ) \frac{\partial \Pi}{\partial r}+\left (T_{{\rm vs}}-\frac{T_{{\rm ev}}T_{{\rm es}}}{T_{{\rm ee}}} \right ) \frac{\partial \mu_{\rm s}}{\partial r} =\frac{r}{4}\frac{\partial z}{\partial t} \qquad \qquad
\IEEEyesnumber \label{eq_313} \IEEEyessubnumber \label{eq_313a}\\
\frac{1}{r} \frac{\partial }{\partial r}\left [ r\left (T_{{\rm vs}}-\frac{T_{{\rm ev}}T_{{\rm es}}}{T_{\rm ee}} \right ) \frac{\partial \Pi}{\partial r}+r\left (T_{{\rm ss}}-\frac{T_{{\rm es}}^2}{T_{{\rm ee}}} \right ) \frac{\partial \mu_{\rm s}}{\partial r} +\frac{r^2n_o(e^{\mu_s/k_BT}-1)}{2}\frac{\partial z}{\partial t} \right ] =\frac{\partial N_s}{\partial t} \qquad \qquad \IEEEyessubnumber \label{eq_313b}
\end{IEEEeqnarray}
Note that the coefficient $T_{ee}$ indeed represents the conductance of the electrolyte film and cannot vanish. 

The quantity $N_s$ and the elements of $\mathsfbi{T}$ are given by the equations (\ref{eq_csce},\ref{eq_chi},\ref{eq_NsNe}) and (\ref{eq_matrix}) as a function of $z(t)$ and $M_s(t)$. Therefore the equations (\ref{eq_313}) 
provide a fully determined set of two non-linear differential equations to be solved with the boundary conditions $\Pi=\mu_\mathrm{s}=0$ at $r=\infty$ and finite second derivative in $r=0$. The pressure is then determined from (\ref{eq_37e}),   giving the electro-hydrodynamic force:
\begin{equation}
F(t)=\int_0^\infty 2\pi r (\Pi+2n(r,t)k_BT[\cosh \chi(r,0,t)-1]dr
\label{eq_315}
\end{equation}

The term $2k_BT n(r,t)[\cosh \chi(r,0,t)-1]$ is the osmotic pressure at the level of the mid-plane. This term decays exponentially with the sphere-plane distance with a screening length equal to the dynamic Debye's length, and represents the direct interaction of the EDL's in dynamic conditions. The term in $\Pi$ corresponds to the long-range hydrodynamic contribution to the force, including all the effects of charge and ions transport described in equations (\ref{eq_313}).


\section{Linear response and contributions to the mechanical impedance} 
\label{sec:linear-response}
\subsection{Linearized transport equations}
At small $h_0$ the above equations are linearized in $h_0$ keeping in mind that terms of order 0 do not vary in time and terms of order 1 are harmonic oscillations of frequency $\omega$.  
In the following, quantities which vanish at equilibrium  such as the fluxes [$J$] and the thermodynamic potentials,  are noted through their complex amplitude as $f(y,r,t)= {\rm Re}[f(r,y)e^{i\omega t}]$, with $\partial / \partial t = i\omega$.
Quantities which do not vanish at equilibrium will be noted $f(r,y,t)=f_{eq}(r,y)+{\rm Re}[{\rm d}f(r,y)e^{i\omega t}]$ with ${\rm d}f \in \mathbb{C}$. An exception will be  the equilibrium film thickness noted $z_{eq}(r)=z(r)=D+r^2/R$ (\ref{approx_parabol}).    

Equations (\ref{eq_313}) are re-written taking into account that:
\begin{enumerate}
    \item \  the thermodynamic potentials $ \Pi$ and   $\mu_{\rm s}$ are of order 1 in $h_0$. We introduce non-dimensional potentials scaled by the amplitude $h_0$ as:
\begin{equation}
m_{\rm v}=\frac{\Pi}{2n_ok_BT}\frac{ \ell_{\rm D}}{h_0} \qquad m_{\rm s}=\frac{\mu_s }{k_BT}\frac{\ell_{\rm D}}{ h_0} \qquad m_{\rm e}=\frac{eW }{k_BT}\frac{ \ell_{\rm D}}{h_0}
\label{eq_41}
\end{equation}
and non-dimensional lengths scaled by $2\ell_D$, such as  $\tilde z=z/2\ell_D$, $\tilde y=y/2\ell_D$ and $\tilde D=D/2\ell_D$, etc... \\
\item \  the term  $(e^{\mu_s/k_BT}-1)\partial z/ \partial t$ in the l.h.s of (\ref{eq_313b}), of order $h_0^2$,  is not kept \\
\item \  in the linearized equations the  transport matrix $\mathsfbi{T}$ restricts to its {\em equilibrium value} $\mathsfbi{T}=\mathsfbi{T}_{eq}$. This means that the $T_{ij}$ have to be calculated with the equilibrium concentration $c_{s,eq}=\cosh \psi_{eq}$ and $c_{e,eq}=-\sinh \psi_{eq}$. A dimensionless transport matrix $\mathsfbi{t} = (t_{ij})$ is introduced as: 
\begin{IEEEeqnarray}{rCl}
t_{ij}=2\int_0^{\tilde z/2}\tilde \Gamma_i(\tilde y)\tilde \Gamma_j(\tilde y) d\tilde y+2\kappa \tilde L_{ij} \qquad i,j \in ({\rm v,s,e)} \qquad 
\IEEEyesnumber \IEEEyessubnumber \\
\kappa=\frac{D_s \pi \eta \ell_B}{k_BT}  \qquad \forall i \quad \tilde L_{{\rm v}i} = 0 
\qquad 
 \IEEEyessubnumber \\
\tilde \Gamma_{\rm v}(\tilde y)=\tilde y \qquad \tilde\Gamma_{\rm s}(\tilde y)=\int_0^{\tilde y} (\cosh \psi_{\rm eq}-1)d\tilde y' \qquad \tilde\Gamma_{\rm e}(\tilde y)=\int_0^{\tilde y} (-\sinh \psi_{\rm eq})d\tilde y' \qquad 
\IEEEyessubnumber \\
\tilde L_{\rm se} =\tilde L_{\rm es}=\tilde \Gamma_e(\frac{\tilde z}{2})+\delta( \tilde \Gamma_s(\frac{\tilde z}{2})+\frac{\tilde z}{2}) \qquad \tilde L_{\rm ss} =\tilde L_{\rm ee}=\tilde \Gamma_s(\frac{\tilde z}{2})+\frac{\tilde z}{2}+\delta \tilde \Gamma_e(\frac{\tilde z}{2})
 \qquad \IEEEyessubnumber 
\end{IEEEeqnarray}
where $\ell_B=e^2/4\pi \epsilon_{l} k_BT$ is the Bjerrum length. In these units the reduced Poiseuille coefficient is $t_\mathrm{vv}=\tilde{z}^3/12$. The non-dimensional diffusivity $\kappa$ is the ratio of the average molecular diffusion coefficient $D_{\rm s}$ to the osmotic diffusion coefficient $K_{DO}=k_BT/\pi \eta \ell_B$ which scales the bulk diffusio-osmotic velocity in presence of a concentration gradient as $v_{{\rm DO}} \propto K_{{\rm DO}} \vec \nabla \ln C$  (\cite{Anderson1989, Siria2013}).
\\
\item \ as $\mathsfbi{T}_{eq}$ (resp. $\mathsfbi{t}$) depends actually on   $z=D+r^2/2R$ (resp. $\tilde z$),  the differential elements in (\ref{eq_313}) are replaced as $r \partial r = Rdz$ and $r/\partial r=2(z-D)/dz$ (resp. 2($\tilde z- \tilde D)/d\tilde z$) \\
\item \ at all places and time the dynamic concentration profile $n_s(r,y,t)$ is equal to the equilibrium profile that would be obtained for the time-dependant values of the film thickness $z(r,t)=z(r)+h_o \cos \omega t$ and of the Debye length $\lambda_D(r,t)=\ell_D e^{-\mu_s/2k_BT}$. 
Therefore the linear response of  $\partial N_s/\partial t$ is a linear expression of $h_0$ and $\mu_s$, noted $dN_s(h_0,\mu_s)$. The associated non-dimensional quantity writes $dN_s / h_0 n_o= a(\tilde z)+b(\tilde z) m_s$ where the non-dimensional functions  $a(\tilde z)$ and $b(\tilde z)$ are expressed in appendix.
\end{enumerate}

Under these circumtances equations (\ref{eq_313}) become two 2nd order linear differential  equations in $\Pi$ and $\mu_{\rm s}$ with non-constant coefficients:
\begin{IEEEeqnarray}{rCl}
\left (T_{\rm vv}-\frac{T_{\rm ev}^2}{T_{\rm ee}} \right ) \frac{d \Pi}{d z}+\left (T_{{\rm vs}}-\frac{T_{{\rm ev}}T_{{\rm es}}}{T_{{\rm ee}}} \right ) \frac{d \mu_{\rm s}}{dz} =\frac{i \omega R h_0}{4} 
  \IEEEyesnumber \label{eq_43} \IEEEyessubnumber  \\
 \frac{d }{dz} \left \{ (z-D)\left [ \left (T_{{\rm vs}}-\frac{T_{{\rm ev}}T_{{\rm es}}}{T_{\rm ee}} \right )  \frac{d \Pi}{d z}+\left (T_{{\rm ss}}-\frac{T_{{\rm es}}^2}{T_{{\rm ee}}} \right ) \frac{d \mu_{\rm s}}{dz} \right ] \right \}=\frac{i\omega R}{2}dN_s
 \IEEEyessubnumber 
\end{IEEEeqnarray}
to be solved for $z\in [D,\infty[$ with boundary conditions of finite second derivative at $z=D$ and zero value at infinity. The non-dimensional counterpart writes:
\begin{IEEEeqnarray}{rCl}
\left (t_{\rm vv}-\frac{t_{\rm ev}^2}{t_{\rm ee}} \right ) \frac{d m_v}{d \tilde z}+\left (t_{{\rm vs}}-\frac{t_{{\rm ev}}t_{{\rm es}}}{t_{{\rm ee}}} \right ) \frac{d m_{\rm s}}{d \tilde z} =\frac{i \omega }{\omega_c} \quad 
  \IEEEyesnumber \label{eq_44} \IEEEyessubnumber  \\
 \frac{d }{d\tilde z} \left \{ (\tilde z- \tilde D)\left [ \left (t_{{\rm vs}}-\frac{t_{{\rm ev}}t_{{\rm es}}}{t_{\rm ee}} \right )  \frac{d m_v}{d \tilde z}+\left (t_{{\rm ss}}-\frac{t_{{\rm es}}^2}{t_{{\rm ee}}} \right ) \frac{d m_{\rm s}}{d \tilde z} \right ] \right \}=\frac{i\omega }{\omega_c }(a(\tilde z)+b(\tilde z)m_s) \quad 
 \IEEEyessubnumber \\
 \omega_c=\frac{2k_BT}{\pi \eta \ell_B R \ell_D} \IEEEyessubnumber \qquad
\end{IEEEeqnarray}

\subsection{Contributions to the mechanical impedance}
The linear force response obtained from eq. (\ref{eq_315})  writes as: 
\begin{eqnarray}
Z=-\frac{F}{h_0} &=& -2\pi R\int_D^\infty \frac{\Pi}{h_0} dz + 4\pi R n_ok_BT(\cosh \psi_{eq,m}(D)-1)  \nonumber \\ 
 & -& 4\pi R n_o \int_D^\infty \frac{\mu_{\rm s}}{h_0} \left (\cosh \psi_{eq,m}-1  +\frac{\partial \cosh \psi_{eq,m}}{\partial (\ln n_o)}  \right ) dz   
\label{eq_42} 
\end{eqnarray}
These terms are splitted in different contributions according to their physical origin. 
\medskip

\begin{itemize}
    \item[-] {\em DLVO stiffness:}
according to (\ref{eq_43}), at zero frequency the fields $\Pi$ and $\mu_{\rm s}$ vanish, and the mechanical impedance resumes to the derivative of the equilibrium DLVO force (\ref{eq_eqforce}):
\end{itemize}
\begin{equation}
    Z(\omega=0)=Z_{\rm DLVO}=-\frac{dF_{\rm DLVO}}{dD} =4 \pi R k_BTn_o(\cosh \psi_{eq,m} (D)-1)
    \label{eq_46}
\end{equation}
This contribution has no imaginary part and is a pure stiffness. 
\medskip

\begin{itemize}
\item[-] {\em Reynolds damping}: at finite frequency the pressure can be splitted in three terms $\Pi_{Rey}+\Pi_{EK}+\Pi_{DK}$ as: 
\end{itemize}
\begin{eqnarray}
\frac{d\Pi_{Rey}}{dz} =\frac{i\omega R h_0}{4T_{vv}}=\frac{6i\omega R h_0 \eta}{z^3}\qquad    \frac{d\Pi_{EK}}{dz} =\frac{i\omega R h_0}{4} \frac{T_{ve}^2}{T_{vv}(T_{vv}T_{ee}-T_{ve}^2)} \qquad \nonumber \\ 
\frac{d\Pi_{DK}}{dz} =-\frac{T_{vs}T_{ee}-T_{ve}T_{se}}{T_{vv}T_{ee}-T_{ve}^2} \frac{d\mu_s}{dz}\qquad 
\end{eqnarray}
The term $\Pi_{Rey}$ gives the so-called Reynolds viscous force exerted by a dielectric liquid drained from the gap between a sphere and a plane: 
\begin{equation}
Z_{Rey}=\frac{6i\pi \eta \omega R^2}{D}
\end{equation}
\medskip

\begin{itemize}
\item[-] {\em Electro-kinetic damping:} the pressure term $\Pi_{EK}$ corresponds to the excess viscous dissipation induced by electrokinetic effects in the electrolyte film when all perturbations in the ions concentration are ignored. Indeed, if one neglects the concentration perturbations in the film, equations (\ref{eq_matrix}) write:
\end{itemize}
\begin{eqnarray*}
eJ_e=0=-T_{\rm ve}\frac{d(\Pi_{Rey}+\Pi_{EK})}{dr}-T_{\rm ee}\frac{dW}{dr} \\ 
J_{\rm v}=-\frac{r}{4}\frac{dz}{dt}=-T_{\rm vv}\frac{d\Pi_{Rey}}{dr}= -T_{\rm vv}\frac{d(\Pi_{Rey}+\Pi_{EK})}{dr}-T_{\rm ve}\frac{dW}{dr}  
\end{eqnarray*}
In other words, the electro-neutrality enforces the streaming current $-T_{\rm ve}d\Pi/dr$ to be compensated by a conductivity current $-T_{\rm ee}dW/dr$, which requires a radial electrical field $-dW/dr$ to develop in the liquid film. This electrical field in return induces an electro-osmotic flow rate $-T_{\rm ve}dW/dr$ which has to be compensated by an additional pressure-driven flow $-T_{\rm vv}d\Pi_{EK}/dr $ in order to maintain the prescribed fluid flow. The electrokinetic effect induces a so-called electro-viscous force, calculated for large distance $D\gg 2\ell_D$ by \cite{Prieve1990}  and also \cite{Rodriguez2022}. Note that this electro-viscous force does not take into account the effect of concentration changes induced by the squeeze flow. 
It  corresponds to a contribution $Z_{EK}$ to the mechanical impedance:
\begin{eqnarray}
 Z_{EK} =\frac{i\omega R}{4}\int_{ D}^\infty \int_{ z}^\infty \frac{T_{\rm ve}^2 \ d z}{T_{\rm vv}(T_{\rm vv}T_{\rm ee}-T_{{\rm ve}}^2)} = \frac{i\omega}{\omega_c}8\pi Rn_ok_BT \mathcal Z_{EK}\nonumber \\
\mathcal Z_{EK}= \int_{\tilde D}^\infty d \tilde z\int_{\tilde z}^\infty \frac{t_{\rm ve}^2}{t_{\rm vv}(t_{\rm vv}t_{\rm ee}-t_{\rm ve}^2)} d\tilde z
\qquad \omega_c=\frac{2k_BT}{\pi \eta \ell_B R \ell_D}
\label{eq_49}
\end{eqnarray}
The electrokinetic impedance $Z_{EK}$ is purely dissipative (imaginary) and strictly proportional to the frequency. For thermodynamic reasons, $T_{\rm vv}T_{\rm ve} \ge T_{\rm ve}^2$ ($t_{\rm vv}t_{\rm ve} \ge t_{\rm ve}^2$), therefore the electrokinetic impedance always corresponds to an excess dissipative force with respect to the nominal Reynolds force. When the fluid conductivity is very large, $T_{\rm ee} = \infty$, the electric field developed in the film becomes negligible, and $Z_{EK}$ vanishes. Section 5 discusses the properties of the real positive non-dimensional factor $\mathcal Z_{EK}$.
\smallskip

\begin{itemize}
\item[-] {\em Diffusio-kinetic impedance:} all remaining contributions involve perturbations of the equilibrium ion concentration and are gathered in an additional impedance that we name here, somewhat abusively, a diffusion-kinetic contribution $Z_{DK}$. We show in appendix that the expression $\cosh \psi_{eq,m}-1 + \partial \cosh \psi_{eq,m} / \partial (\ln n_o)$ appearing in equation (\ref{eq_42}) resumes to the function $a(\tilde z)$, so that $Z_{DK}$ writes as:
\end{itemize}
\begin{eqnarray}
  Z_{DK}=8\pi R n_ok_BT \mathcal Z_{\rm DK}\qquad  \nonumber\\
  \mathcal Z_{\rm DK}=-\int_{\tilde D}^\infty d\tilde z \left ( a(\tilde z) m_s(\tilde z)   +   \int_{\tilde z}^\infty g(\tilde z') \frac{dm_s(\tilde z')}{d\tilde z'}d\tilde z' \right ) \qquad g(\tilde z)= \frac{t_{\rm vs}t_{\rm ee}-t_{\rm ve}t_{\rm es}}{t_{\rm vv}t_{\rm ee}-t_{\rm ve}^2} \qquad 
\label{eq_410}
\end{eqnarray}
where the reduced potential $m_s$ is solution of the 2nd order linear differential equation obtained by decoupling the two equations (\ref{eq_44}):
\begin{IEEEeqnarray}{rCl}
 \frac{d}{d\tilde z}\left \{ (\tilde z - \tilde D)\left (h(\tilde z) \frac{dm_{\rm s}}{d\tilde z}+\frac{i\omega}{\omega_c}g(\tilde z) \right ) \right \} = \frac{i\omega}{\omega_c} \left (a(\tilde z)+b(\tilde z)m_{\rm s} \right)
\qquad  \IEEEyesnumber \label{eq_48} \IEEEyessubnumber \\
 h(\tilde z)=t_{\rm ss}-\frac{t_{\rm se}^2}{t_{\rm ee}}- \frac{(t_{\rm vs}-t_{\rm ve}t_{\rm se}/t_{\rm ee})^2}{t_{\rm vv}-t_{\rm ve}^2/t_{\rm ee}} \qquad
 \omega_c=\frac{2k_BT}{\pi \eta \ell_B R \ell_D} \qquad 
\IEEEyessubnumber 
\end{IEEEeqnarray}
The first term of $\mathcal Z_{DK}$ appearing in equation (\ref{eq_410}) accounts for the modification of the disjonction pressure in the middle-plane due to the perturbed ion concentration. This term is short-ranged as the function $a(\tilde z)$ vanishes exponentially with $\tilde z=z/2l_D$. It represents the out-of-equilibrium direct interaction of the EDL's. The second term gathers all the transport effects induced by the concentration gradient $\partial \mu_s/\partial r$. The concentration gradient induces a diffusio-osmotic flow, which in the same way as the electro-osmotic flow, has to be compensated by an additional Poiseuille flow in order to maintain the prescribed flow rate in the liquid film. Furthermore diffusio-osmotic effects contribute to a diffusio-osmotic charge current, which in turn modify the electro-kinetic charge restauration effect in the electrolyte film. These contributions are embeded in the terms $t_{\rm vs}t_{\rm ee}/(t_{\rm vv}t_{\rm ee}-t_{\rm ve}^2)$ and $t_{\rm ve}t_{\rm se}/(t_{\rm vv}t_{\rm ee}-t_{\rm ve}^2)$ of the $g$ function.

As ions can accumulate, $Z_{DK}$ is  frequency-dependant with an imaginary as well as a real component. It thus contributes both to the stiffness and to the damping of the global impedance. The frequency behaviour or $Z_{DK}$ is scaled by the system frequency $\omega_c$, associated to the osmotic diffusion time (with diffusion coefficient $K_{DO}=k_BT/\pi \eta \ell_B$) over the lateral extension $\sqrt{R \ell_D}$ of the EDL's contact.

\subsection{Summary}
In summary our theory describes the mechanical impedance in a drainage flow of a monovalent  electrolyte as a function of the normalized distance $\tilde D=D/2\ell_D$ and 4 non-dimensional parameters: 
\begin{itemize}
\item[$\bullet \ $] the parameter $\alpha$ characterizes the electrostatic properties of the surfaces in the solution of concentration $n_o$. More specifically $\alpha$ is related to the nominal surface potential $V_{s,nominal}$ of the surfaces alone (non-interacting)  by 
$\displaystyle \alpha = 2 \sinh (eV_{s,nominal}/2k_BT)$. In water at 300 K, the value $\alpha=1$ corresponds to a nominal surface potential of 24 mV, $\alpha=10$ to 115 mV, $\alpha=100$ to 230 mV. 
\item[$\bullet \ $] the parameter $\kappa$ characterizes the average diffusion coefficient of the electrolyte. If anions and cations have a similar diffusion coefficient, $\kappa$ is related to the hydrodynamic radius of the ions defined by the Stokes-Einstein relation $D_s=k_BT/6\pi \eta r_H$ by \hbox{$\kappa= \ell_B/6 r_H$} where $\ell_B$ is the Bjerrum length. For the KCl electrolyte of diffusion coefficient \hbox{1.45 $10^{-9}{\rm m}^2/$s} in water at 300K, $\kappa=0.7$.  
\item[$\bullet \ $] the parameter $\delta$ describes the diffusion contrast between co-ions and the counter-ions. The diffusion coefficient of counter-ions is $D_s(1-\delta)$.
\item[$\bullet \ $] the frequency ratio $\beta=\omega/\omega_c$. It is of interest to note that the combination \hbox{$\beta/\kappa =\omega/\kappa \omega_c= \omega R\ell_D/2D_s$} characterizes the ratio of the diffusion time over the lateral extension of the overlapping EDL's at contact $D=0$, to the period of the oscillation. 
\end{itemize}
The dimensional real and imaginary part of the mechanical impedance write:
\begin{IEEEeqnarray}{rCl}
\Re Z=8\pi R n_ok_BT\left (\frac{1}{2}[\cosh \psi_{m,eq}(\tilde D,\alpha)-1]+\Re \mathcal Z_{DK}(\tilde D, \alpha,\beta,\kappa,\delta) \right ) \IEEEyesnumber \label{eq_412} \IEEEyessubnumber\\
\Im Z= \frac{6 \pi \omega \eta R^2}{D}(1+\tilde f_{EK}+\tilde f_{DK} )
\IEEEyessubnumber\\
\tilde f_{EK}(\tilde D,\alpha,\kappa,\delta)=\frac{\tilde D }{6}\mathcal Z_{EK} \qquad \tilde f_{DK}(\tilde D,\alpha,\beta,\kappa,\delta) = \frac{\tilde D }{6 \beta} \Im \mathcal Z_{DK} \label{eq_412c} \IEEEyessubnumber
\end{IEEEeqnarray}


\section{Properties of the  electro-kinetic damping}
In the following the non-dimensional electro-kinetic factor $\mathcal Z_{EK}$ is calculated by integrating numerically equation \ref{eq_49}, the transport coefficients being themselves calculated by numerical integration of  equations \ref{eq_42} with the expressions of $\tilde \Gamma_{e,s}$ given in Appendix. Details of implementation are given in the Supplementary Informations. We find that at $\tilde D \ge 10$ the transport coefficients meet fully the large distance expressions given in Appendix D, and we use these expressions. 

\subsection{Large distance behaviour.}
Electrostatic Double Layers are considered  non-overlapping if the distance between the surfaces is very large compared to the Debye's length, $D\gg 2\ell_D$. In this case the equilibrium  potentiel in the mid-plane vanishes, $\psi_{eq,m}=0$, and the EDL's do not interact at equilibrium. The coefficients of the transport matrix $\mathsfbi{t}$, at the exception of $t_{{\rm vv}}=\tilde z^3/12$, are all affine functions of $\tilde z$ and write as the sum of a bulk part proportional to $\tilde z$ and of a surface part:  $t_{ij}=a_{ij} \tilde{z}+b_{ij}$ (see Appendix). More specifically the bulk part of  $t_{{\rm ee}}$  is due to molecular diffusion only, while the bulk part of $t_{{\rm ve}}$ corresponds  to the bulk electro-osmotic flow and depends only on the surface potential $\psi_s$:
\begin{equation}
   a_{{\rm ee}} =\kappa 
    \qquad  a_{{\rm ve}}=\frac{\psi_s}{4}=\frac{1}{2} {\rm asinh} \frac{\alpha}{2}
\end{equation}
The surface parts $b_{\rm ee}$ and $b_{\rm ve}$ correspond to transports in the EDL's and their detailed expression as a function of $\kappa$ and $\alpha$ is given in Appendix. 

In these conditions the function $t_{\rm ve}^2/t_{\rm vv}(t_{\rm vv}t_{\rm ee}-t_{\rm ve}^2)$ to integrate in order to get $\mathcal Z_{EK}$ is a rational fraction whose denominator is an always positive  fourth degree polynomial. The non-dimensional EK damping writes:
\begin{equation}
\mathcal Z_{EK}=\frac{144}{a_{\rm ee}}\sum_{i=1}^4\frac{A_i}{a_i^2}\left [ \left (\frac{\tilde D}{a_i}-1 \right ) \ln \left (1-\frac{a_i}{\tilde D} \right ) +1-\frac{a_i}{2\tilde D}\right ]
\label{eq_52}
\end{equation}
where the fonction ln is the complex logarithm, the $a_i$'s are the four complex roots of $t_{\rm vv}t_{\rm ee}-t_{\rm ve}^2$, and the complex coefficients $A_i$'s obey the relations:
$$A_i=\frac{(a_{\rm ve}a_i+b_{\rm ve})^2}{\prod_{j\neq i}(a_i-a_j)} \qquad \sum_{i=1}^4 A_i=\sum_{i=1}^4 \frac{A_i}{a_i^2}=\sum_{i=1}^4 \frac{A_i}{a_i^3}=0 \quad \sum A_ia_i=a_{\rm ve}^2  
$$
We find numerically that the analytical expression (\ref{eq_52}) holds up to $\tilde D \sim 4$. At very large distance the leading term of $\mathcal Z_{EK}$ is in $1/D^3$ as first calculated by \cite{Prieve1990}:
\begin{equation}
    \mathcal Z_{EK} \simeq \frac{12 a_{\rm ev}^2}{a_{\rm ee}\tilde D^3} =\frac{3 \psi_s^2}{4 \kappa \tilde D^3} 
    \qquad Z_{EK}=i\omega \frac{R^2 \ell_D^2}{\ell_B D^3} \frac{3 eV_s}{D_s}
    \label{eq_53}
\end{equation}
\begin{figure}
  \centering
  \includegraphics[width=0.7\linewidth]{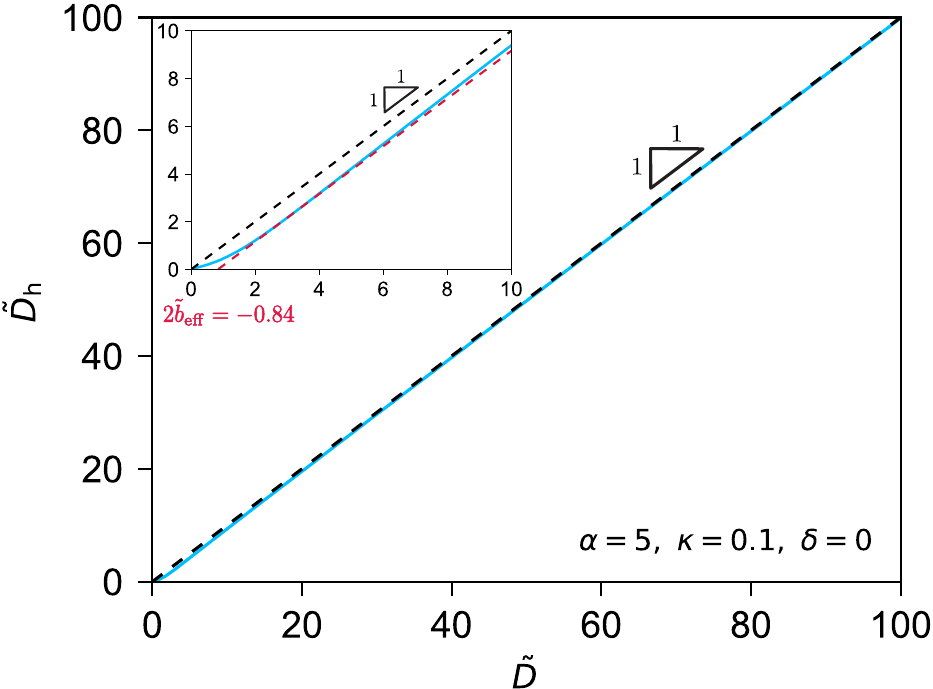}  
  \caption{Normalized hydrodynamic distance $\tilde D_h=D_h/2\ell_D = \tilde D/(1+\mathcal Z_{EK}/\mathcal Z_{Reynolds}) $ as a function of $\tilde D$, for $\alpha=5$, $\kappa=0.1$ and $\delta = 0$. The dashed line is the diagonal $\tilde D_h=\tilde D$. Inset: expansion in the distance range $\tilde D \le 10$. The red dashed line corresponds to $\tilde D_h=\tilde D+2 b_{eff}/2\ell_D$ with $b_{eff}=-0.84 \ell_D$.}
\label{fig:51}
\end{figure}

Experimentally it can be of interest to characterize the excess damping due to the electro-kinetic   effect by an hydrodynamic distance between the surfaces, defined as $D_h=6\pi \eta \omega R^2/\Im(Z)$. Indeed when the damping reduces to the Reynolds damping, $D_h$ is the distance separating the no-slip planes. In presence of an excess damping, the hydrodynamic distance reflects the virtual position of the solid surfaces that would produce an equivalent Reynolds damping, in other words it accounts for the  excess dissipation by (apparent) stagnant layers at wall of thickness $b_{app}=(D-D_h)/2$. Figure \ref{fig:51} shows a typical plot of the hydrodynamic distance 
versus the real distance,  showing actually $D_h<D$. In a range of distance of some tens of Debye's length, the difference of $D-D_h$   is non-negligible and corresponds to apparent stagnant layers of the order of the Debye length. 
However these apparent stagnant layers  are not constant, their thickness decreases and decays to zero when $\tilde D \rightarrow \infty$. This is in agreement with  the expansion of $\mathcal Z_{EK}$  at large $\tilde D$ which has no $\tilde D^{-2}$ term  ($(\tilde D-2b_{app})^{-1} \sim \tilde D^{-1}+2b_{app} \tilde D^{-2} $ at $\tilde D \rightarrow \infty$). Thus the electro-kinetic effect, although it induces an increased dissipative mechanism whose origin lies in the surface charge and the EDL's,  cannot be described by a well-defined and non-ambiguous hydrodynamic distance based on apparent stagnant fluid layers at the solid walls.

\subsection{Small and intermediate distances.}
\begin{figure}
  \centering
  \includegraphics[width=0.49\linewidth]{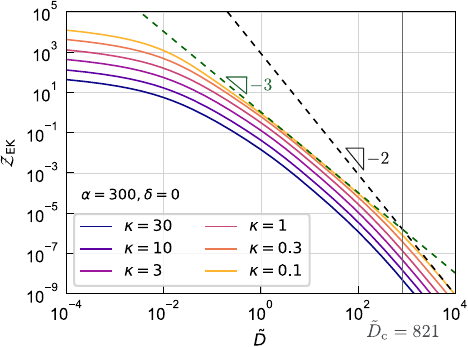}  
  \includegraphics[width=0.49\linewidth]{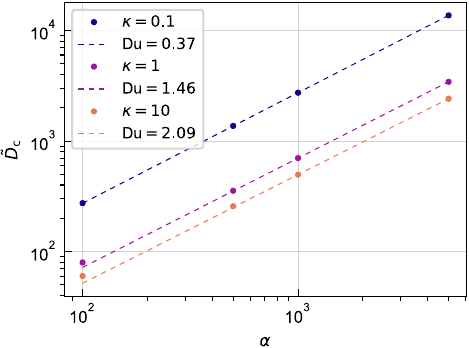}  
  \caption{Left: Adimensional electrokinetic damping $\mathcal Z_{EK}$ as a function of the normalized distance $\tilde D=D/2\ell_D$ for the parameter $\alpha=300$, $\delta=0$, and various values of $\kappa$. The dashed line of slope -3 plots the scaling law (\ref{eq_53}), the dashed line of slope -2 plots the scaling law (\ref{eq_55}). Right: crossing distance $\tilde D_c$ between the power laws in $\tilde D^{-2}$ and in $\tilde D^{-3}$  followed by $\mathcal Z_{EK}$, as a function of $\alpha$. The dashed lines correspond to the values of the Dushin number given by equation \ref{eq_56}.
  }
\label{fig:52}
\end{figure}

At $\tilde D<1$  the EDL's overlap. One observes on figure \ref{fig:52} a quasi-saturation of $\mathcal Z_{EK}$.  In this case a simplified model for the thin electrolyte film is the so-called Donnan's model, explicited in Appendix E. The electro-kinetic factor becomes to the leading order: 
\begin{equation}
\frac{t_{\rm ve}^2}{t_{\rm vv}(t_{\rm vv}t_{\rm e}-t_{\rm ve}^2)}\simeq \frac{\alpha}{\tilde z^2 \kappa(1-\delta)} \qquad \mathcal Z_{EK}\simeq -\frac{\alpha \ln \tilde D}{\kappa(1-\delta)}
\label{eq_54}
\end{equation} 

At $\tilde D >1$, figure \ref{fig:52} shows that an intermediate  behaviour develops at high enough surface charge, not following the $\tilde D^{-3}$ asymtotic power law (\ref{eq_53}), but rather a $D^{-2}$ power law. In this distance range the EDL's do not overlap ; but this intermediate behaviour is due to the fact that  the electrical conduction of the electrolyte film is dominated by the counter-ions of the EDL's rather than the bulk solution: $t_{\rm ee} \approx b_{\rm ee}$. This occurs when $\tilde z<  \vert b_{\rm ee}/a_{\rm ee}\vert = \alpha[\gamma(1+1/2\kappa) - \delta]$. In this range of distance the electrokinetic damping writes: 
\begin{equation}
\mathcal Z_{EK} \simeq \int_{\tilde D}^\infty d\tilde z\int_{\tilde z}^\infty \frac{144 a_{\rm ve}^2}{b_{\rm ee}^2\tilde z'^6}d\tilde z' = \frac{24 a_{\rm ve}^2}{b_{\rm ee} \tilde D^2}
\label{eq_55}
\end{equation}
This scaling law is fully supported by the data as shown in figure \ref{fig:52}.  

The dominance of the EDL's conduction in a film of thickness $h$ is generally characterized by the Dushin number Du=$\sigma/2 n h$. In our case the Dushin number at thickness $z$ is Du=$\alpha/\tilde z$. The crossing distance between the intermediate regime described by the law (\ref{eq_55}) and the large distance power law (\ref{eq_53}) is indeed expected to occur at $\tilde D_c=b_{\rm ee}/2a_{\rm ee}$, that is a Dushin number  Du$_c=2\alpha a_{\rm ee}/ b_{\rm ee}$, expressing as (see Appendix D): 
\begin{eqnarray}
    \tilde D_c = \frac{\alpha}{2}[\gamma(1+\frac{1}{2\kappa})-\delta] = \frac{\alpha}{{\rm Du}_c } \qquad {\rm Du}_c =2[\gamma+\frac{\gamma}{2\kappa}-\delta]^{-1}
    \label{eq_56}
\end{eqnarray}
The comparison of this law to the cross-over distance determined from the numerical data in figure \ref{fig:52} right, supports fully the  interpretation of the intermediate range of $\mathcal Z_{EK}$ in terms of Dushin number and dominance of the electrical conduction of the EDL's.

\subsection{Comparison to the Reynolds damping}
We use  the relative amplitude factor $\tilde f_{EK}=(\tilde D/6) \mathcal Z_{EK}$ defined in equation \ref{eq_412c} to compare the electro-kinetic damping to the Reynolds damping. 

\begin{figure}
  \centering
  \includegraphics[width=0.49\linewidth]{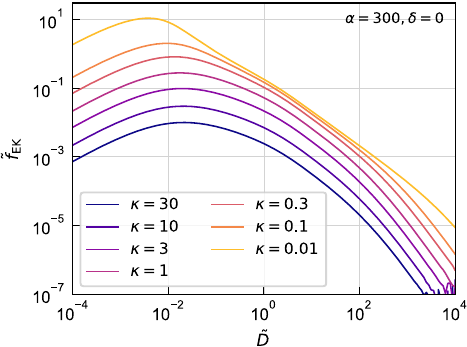}  
    \includegraphics[width=0.49\linewidth]{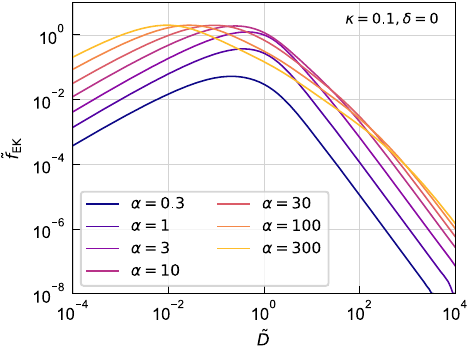}  
      \includegraphics[width=0.6\linewidth]{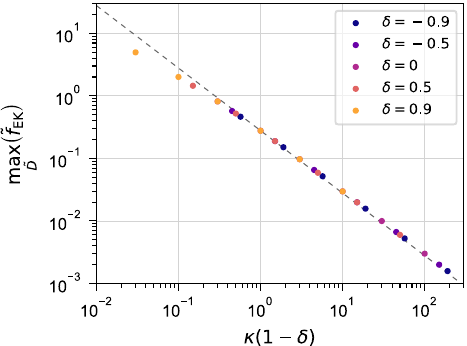}  
  \caption{Top left: relative electro-kinetic
damping $\tilde f_{EK}=\mathcal Z_{EK}\tilde D/6$ as a function of the normalized distance $\tilde D=D/2\ell_D$ for $\alpha=300$, $\delta=0$, and various values of $\kappa$. Top right: $\tilde f_{EK}$ as a function of $\tilde D$ for $\kappa=0.1$,   $\delta=0$, and various values of $\alpha$. Bottom: maximum value of the relative electrokinetic damping $\tilde f_{EK}$, for $\alpha > 50$, as a function of the normalized diffusion coefficient of the counter-ions $\kappa(1-\delta)$. Values of the diffusion contrast $\delta$ are reported in the legend. Values of $\kappa $ can be read on the x-axis at $\delta=0$, and are (0.3,1,3,10,30,100). The maximum value of $\tilde f_{EK}$ does not depend on $\alpha$.}
\label{fig:53}
\end{figure}

Figure \ref{fig:53} top-left shows the influence of the diffusion parameter $\kappa$ on $\tilde f_{EK}$. Without surprise $\tilde f_{EK}$ decreases when $\kappa$ increases, as this corresponds to an increase of the electrolyte conductivity, and therefore a decrease of the electric field required to restore the global electro-neutrality.  In the regions of very strong overlap ($\tilde f_{EK} \sim D \ln D$) and of non-overlap with small Dushin ($\tilde f_{EK} \sim 1/D^2$)  $\tilde f_{EK}$ decreases essentially in $1/\kappa$. In the intermediate distance range however, $\tilde f_{EK}$ saturates to a diffusion-independant limit at low $\kappa$. This effect is due to the main contribution of the electro-osmotic flow to the electrolyte conductivity  in this range of film thickness at low diffusivity and large surface charge. 

Figure \ref{fig:53} top-left shows the influence of the electrical parameter $\alpha$ on $\tilde f_{EK}$. At small values of $\alpha$ the relative amplitude of the electro-kinetic damping increases with the surface charge, as expected intuitively. However above $\alpha \ge 10$ (for the particular value of $\kappa=0.1$), the maximum value reached by $\tilde f_{EK}$ becomes totally independent of $\alpha$. The only effect of increasing $\alpha$ is that the maximum of $\tilde f_{EK}$ is reached for lower and lower distances. This effect is in fact associated to a non-monotonic variation of $\tilde f_{EK}$ with $\alpha$ at a given distance $\tilde D$. For $\alpha$ values large enough so that the intermediate distance range of high Dushin number (with $\tilde f_{EK} \sim D^{-1}$) develops, $\tilde f_{EK} $ is a decreasing function of $\alpha $. The amplitude of $\tilde f_{EK}$ in this intermediate distance range is indeed determined by the ratio $a_{\rm ve}^2/b_{ee}$ (see equation \ref{eq_55}), which is in fact the square of the amplitude of the bulk electro-osmotic flow, divided by the contribution of this flow to the electrical conductivity. Clearly the bulk electro-osmotic flow grows as $\psi_{s, nominal} \sim \ln \alpha$, whereas its contribution to the conductivity grows as $\alpha \gamma \sim \alpha$ at high surface charge (see Appendix D for the expression of $b_{\rm ee}$). Therefore, the increase of conductivity dominates in the intermediate regime, and induces the inversion of growth of $\tilde f_{EK}$ with $\alpha$.

Finally the overall maximum value of the relative electro-kinetic damping is plotted in figure \ref{fig:53} bottom for values of $\alpha \ge 50$. A remarkable variation  is found as the inverse of the normalized diffusion coefficient of the counter ions, $\kappa (1-\delta)$. This behaviour  reflects the fact that the properties of overlapping EDL's are determined by the ones of the counter-ions. Electro-kinetic dampings may reach 100\% of the Reynolds damping for $\kappa (1-\delta) \le 0.3$, which means counter-ions of hydrodynamic radius slightly less than 0.4 nm (in water at 300K). This property of the maximum relative electro-kinetic damping could provide a method for measuring quite accurately the diffusion coefficient of single ionic species.


\section{Properties of the  diffusio-kinetic impedance}
The diffusio-kinetic impedance is calculated by solving numerically the linear differential equation (\ref{eq_48}) and integrating equation (\ref{eq_410}). These operations are performed all-at-once using Python ODE solving routines, details of implementation and codes are reported in the Supplementary Materials. 

\subsection{Diffusio-kinetic stiffness}
The diffusio-osmotic stiffness $\Re Z_{DK}$ takes its origin in ions accumulation, as described by the r.h.s of equation (\ref{}). 

\smallskip
{\em \parindent=0pt High frequency upper
limit at $ \omega \ge \omega_c \kappa$.}
\smallskip

At high frequency the ions  follow adiabatically the motion of the sphere: the perturbation in ion concentration is fully in phase with the displacement, and both the modification of the disjonction pressure as well as the diffusio-osmotic flow are  in phase with the displacement.  The additional pressure $\Pi_{DK}$ needed to counter-balance the diffusio-osmotic flow is thus also in phase with the displacement. 
In this high frequency   limit the diffusio-kinetic stiffness is frequency-independant and takes the analytic form:
\begin{IEEEeqnarray}{rCl}
    \Re m_s^{hf} =\frac{g+g'(z-\tilde D)-a(\tilde z)}{b(\tilde z)} \IEEEyesnumber \IEEEyessubnumber \\
    \Re \mathcal Z_{DK}^{hf} = \int_{\tilde D}^\infty d\tilde z\left ((g-a)\Re m_s^{hf} + \int_{\tilde z}^\infty g'\Re m_s^{hf} d\tilde z' \right )
\IEEEyessubnumber
\label{eq_61} \end{IEEEeqnarray}
which is calculated by a simple integration (see Sup. Mat). 

\begin{figure}
        \includegraphics[width=0.49\textwidth]{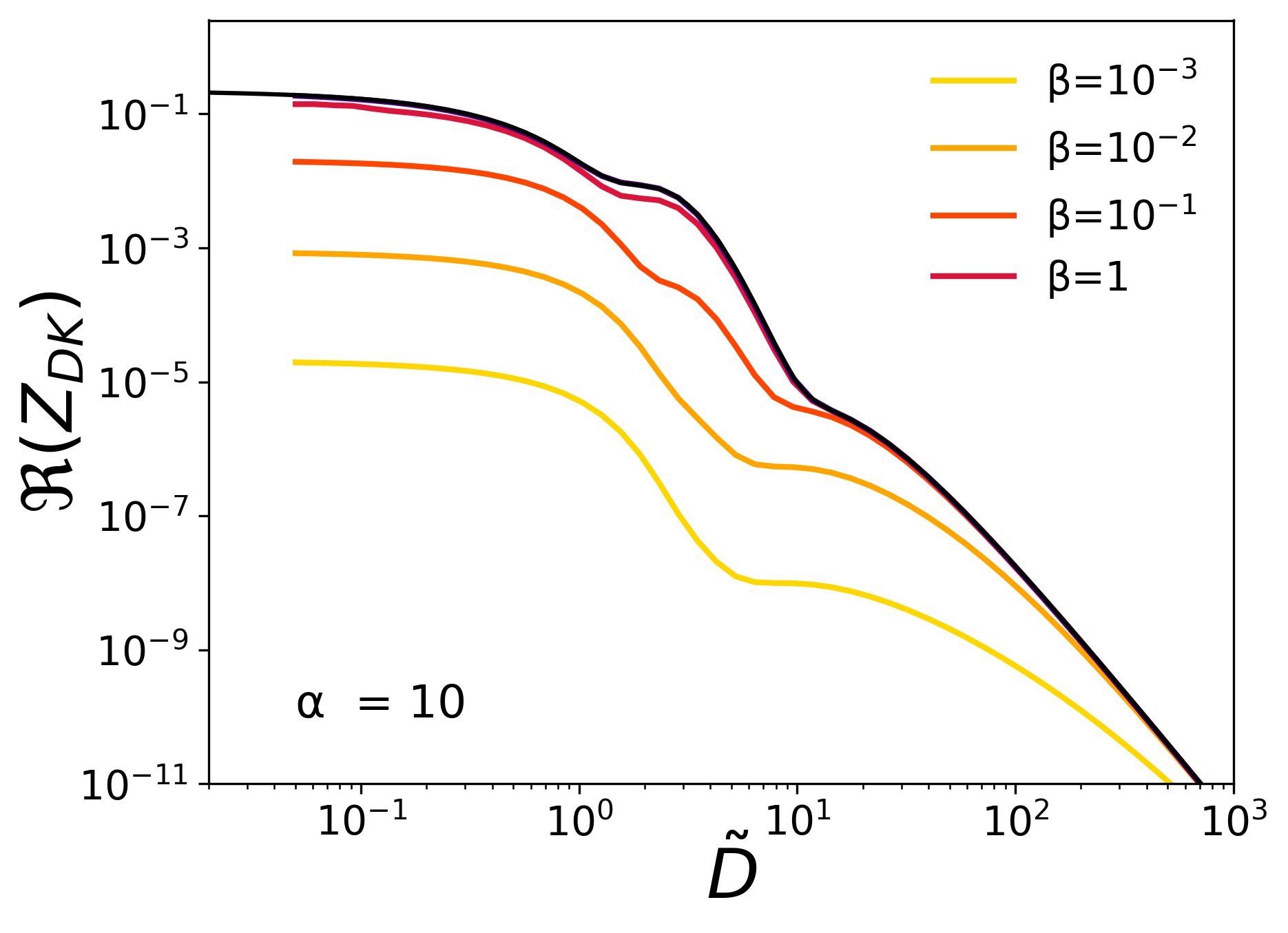}
       \includegraphics[width=0.49\textwidth]{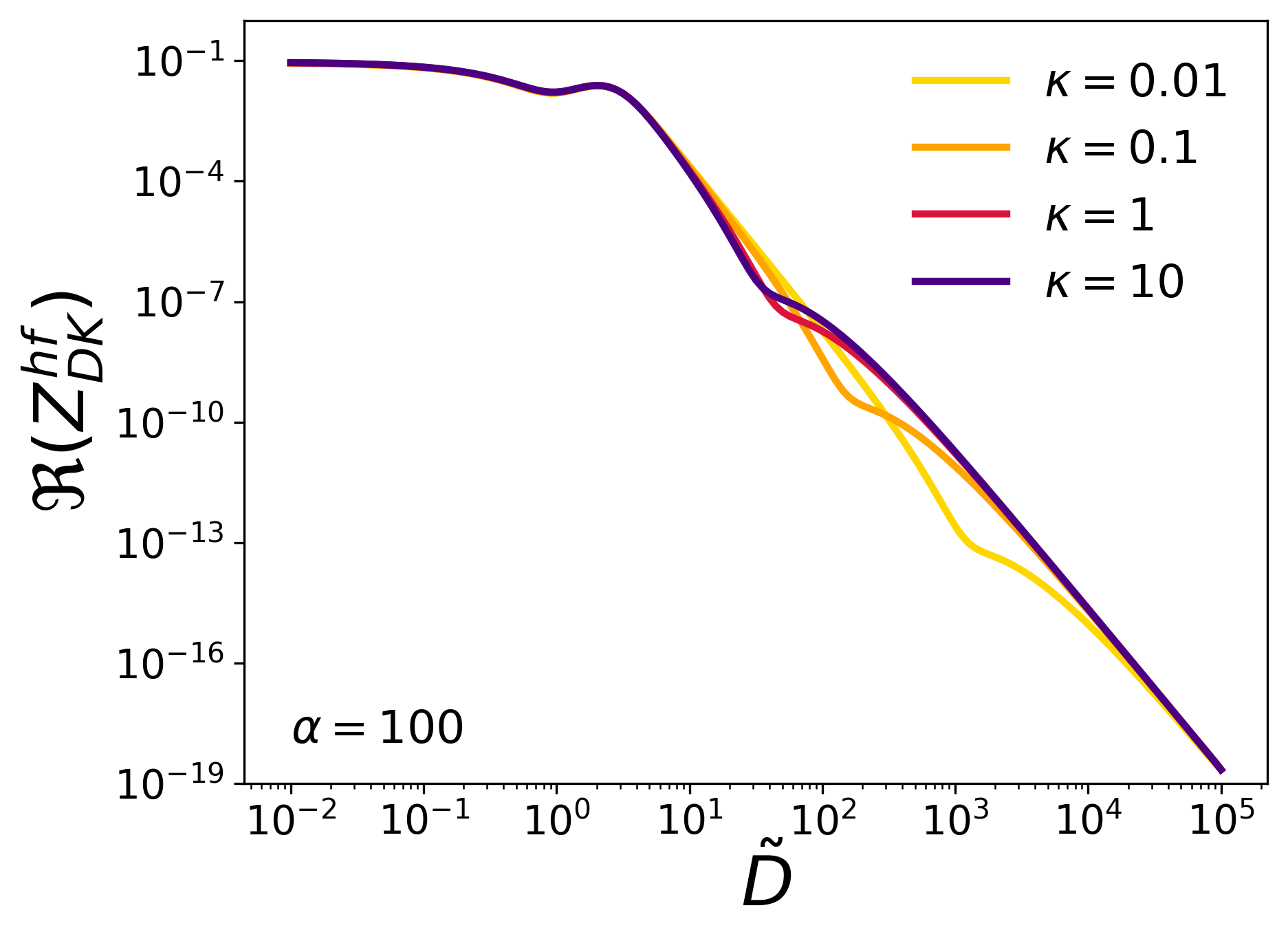}
    \caption{Left: real part of the diffusio-kinetic impedance $\Re(\mathcal Z_{DK})$ as a function of the adimensional gap $\tilde D=D/2\ell_D$, for $\alpha=10$, $\kappa=1$ and various values of $\beta=\omega/\omega_c$. The black line correspond to the high frequency limit eq. (\ref{eq_61}). Right:  high frequency stiffness $\Re(\mathcal Z_{DK}^{hf})$ as a function of $\tilde D$ for $\alpha=100$ and various values of $\kappa$.
    }
    \label{fig:61}
\end{figure}

 Figure \ref{fig:61} left  shows a typical  plot of the variation of $\Re(\mathcal Z_{DK})$ as a function of $\tilde D$ for $(\alpha,\kappa)=(10,1)$ and various values of $\beta=\omega / \omega_c $. We observe that for all the distance range,  the high frequency behaviour is an upper limit for the diffusio-kinetic stiffness: $\Re \mathcal Z_{DK} \le \Re \mathcal Z_{DK}^{hf}$. This property holds for all parameters $\alpha$ and $\kappa$, therefore the highest possible values of the stiffness are always obtained when the convergence to the high-frequency limit is reached.

 In practice we observe that the convergence is essentially reached when   $ \omega / \omega_c \kappa  \ge 1$. This appears on figure \ref{fig:61} left for the particular value $(\alpha,\kappa)=(10,1)$, and tests with various values of $(\alpha,\kappa)$ fully support this property. The time $(\omega_c \kappa)^{-1} =R\ell_D/2D_s $ indeed reprensents the molecular diffusion time over the lateral extension $\sqrt{2R\ell_D}$ of the overlapping EDL's when $\tilde D=0$. This time is the shortest system's time found on the route toward restauring the equilibrium, therefore when it is larger than the oscillation period, the high frequency limit is reached.  

 Finally the convergence toward the high frequency limit is not uniform over the distance range, and occurs for lower frequencies at large distances. More specifically the convergence at distance $\tilde D$ is essentially obtained for $\omega \simeq  D_s/R \ell_D \tilde D $, which is the inverse of the diffusion time over the lateral extension of the hydrodynamic contact $\sqrt{2RD}$.

\smallskip
{\em \parindent=0pt Long-range decay and comparison with the DLVO stiffness}
\smallskip

At large distance when EDL's do not overlap, $dN_s \simeq 2z(n-n_o)$, thus $a(\tilde z) \simeq 0$ and $b(\tilde z) \simeq 2 \tilde z$. The transport coefficients $t_{ij}$ are affine functions of $\tilde z$ except for $t_{\rm vv}= \tilde z^3/12$, and $g(\tilde z) $ resume to the leading order to $ g_2/\tilde z^2$, $g_2=12 a_{\rm vs}$ if $\delta = 0$. The high-frequency stiffness has the asymptotic form:
\begin{equation}
 \Re \mathcal Z_{DK}^{hf} \sim \frac{42 a_{\rm vs}^2}{5\tilde D^4}
 \qquad a_{\rm vs}=-\frac{1}{2} \ln (1-\gamma^2)
 \label{eq_62}
\end{equation}
In this large distance regime, the long range decay of the stiffness in $1/D^4$ is fully due to the transport of the excess ions concentration in the EDL's by the Poiseuille flow and the resulting diffusio-osmotic flow thereby induced. The long range decay depends only on the surface charge $\alpha$ and does not involve any diffusion process. 
\begin{figure}
        \includegraphics[width=0.75\textwidth]{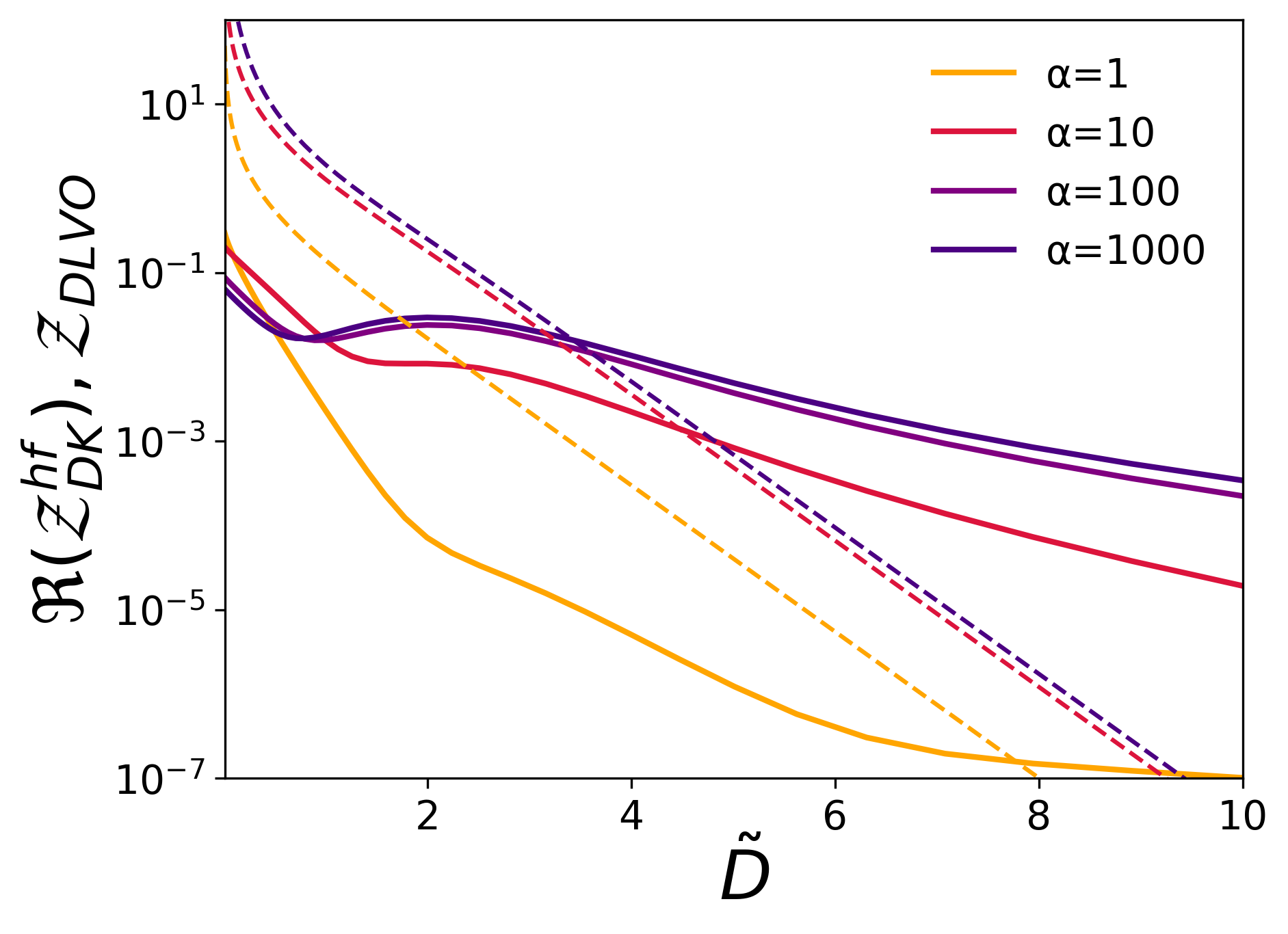}      
    \caption{Plain lines: non-dimensional high-frequency diffusio-kinetic stiffness $\Re \mathcal Z_{DK}^{hf}$ for various values of $\alpha$. The diffusion contrast $\delta$ is set to 0. Dashed lines: non-dimensional DLVO stiffness $\mathcal Z_{DLVO}=Z_{DLVO}/8\pi R n_ok_BT$ for $\alpha=1$ (yellow), $\alpha=10$ (red), and $\alpha=100$ (indigo). For higher values of $\alpha$ the DLVO stiffness is essentially the same as for $\alpha=100$.
    }
    \label{fig:62}
\end{figure}

At short distance the stiffness saturates (see  figure \ref{fig:61} right)  to a value which is also independent of diffusion.  As a whole the influence of diffusion  in the high-frequency stiffness appears only in the intermediate range $1 < \tilde D < \alpha$, that is when  the EDL's do not overlap but the Dushin number is larger than 1. In this intermediate range the excess ions in the EDL's play a more important role in the transport properties than the bulk, resulting in an interplay between convective and diffusive transport. Note that the stiffness decays quite fast and only the short distance behaviour up to $\tilde D=10$ might fall in an experimental range.

It is of interest to compare the diffusio-kinetic stiffness to  the DLVO stiffness, which for $\tilde D>1$ decreases exponentially with $\tilde D$. As the dimensional prefactor is in $R n_o k/BT$ for both quantities, this comparison can be made on the non-dimensional expressions of the stiffness, and is independent on the sphere radius. One sees on figure (\ref{fig:62}) that at distance sufficiently large the diffusio-kinetic stiffness always exceeds the DLVO stiffness. Furthermore the inversion of magnitude occurs essentially at a distance where the high-frequency diffusio-kinetic  stiffness does not depend on  $\kappa$. For the $\alpha=1$ the crossing distance is about $\tilde D=8$, at a very low value of the stiffness, and should not be possible to reach experimentally. However for higher values of $\alpha$ the crossing distance decreases and the stiffness values increases significantly, reaching  somewhat typical distance values $\tilde D \sim 3.5 - 4$ and stiffness value $ \sim 0.05\times 8\pi Rn_ok_BT$ at $\alpha >20$.

\subsection{Diffusio-kinetic damping}

\smallskip
{\em \parindent=0pt Variations with frequency.}
\smallskip

\begin{figure}
           \includegraphics[width=0.49\textwidth]{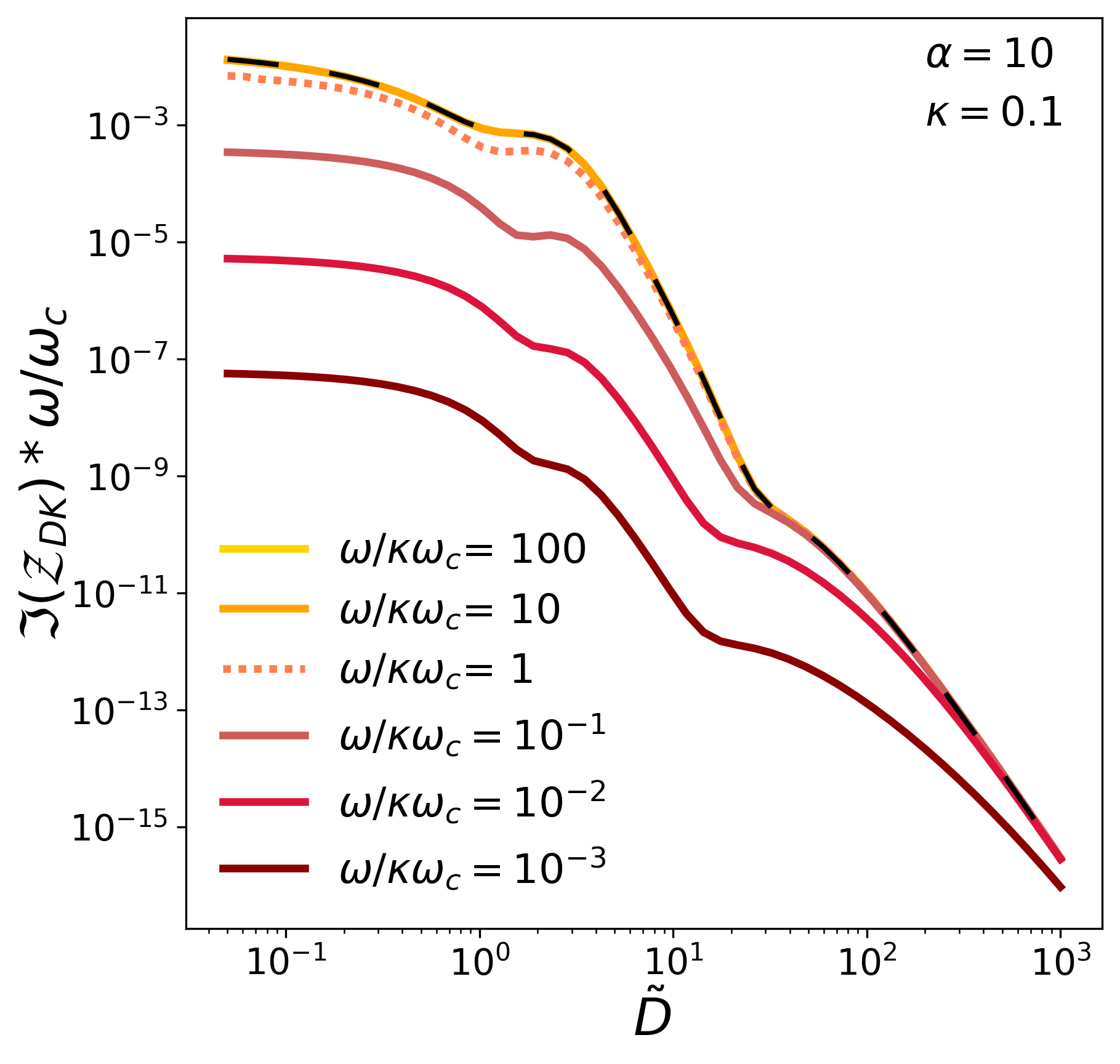}
       \includegraphics[width=0.49\textwidth]{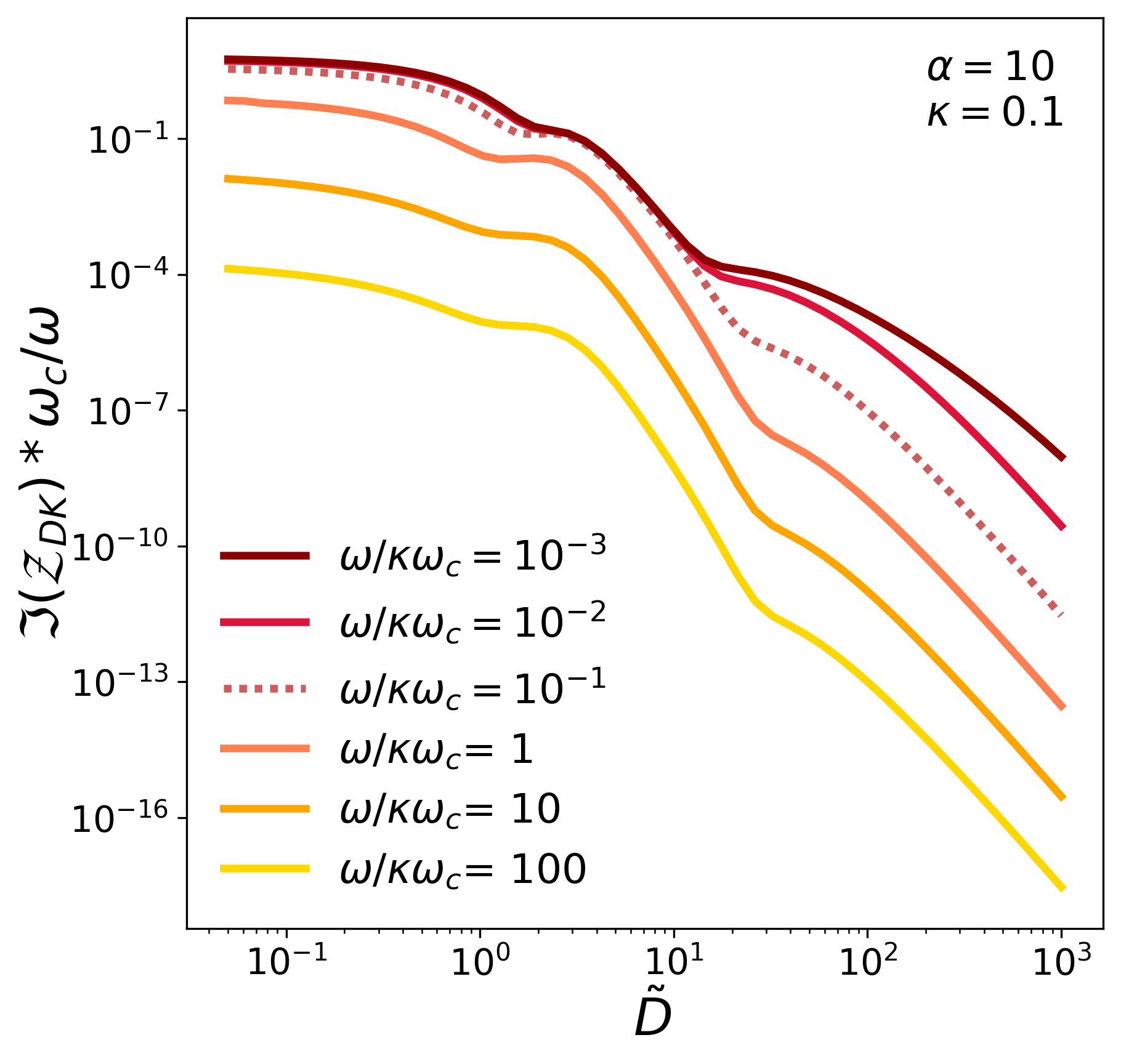}
    \caption{Left: imaginary part of the diffusio-kinetic impedance multiplied by the frequency ratio $\Im(\mathcal Z_{DK})*\omega/\omega_c$  as a function of the adimensional gap $\tilde D=D/2\ell_D$, for $\alpha=10$, $\kappa=0.1$ and various values of $\omega/\kappa \omega_c$. For values $\omega/\kappa \omega_c\ge 1$ all the data are very close to,  or collapse on the black dashed line. Right: imaginary part of the diffusio-kinetic impedance divided by the frequency ratio  $\Im(\mathcal Z_{DK})*\omega_c/\omega$ as a function of $\tilde D$ for the same values of $\alpha$, $\kappa$ and $\omega/\kappa \omega_c$ as on the left graph. For values $\omega/\kappa \omega_c\le 10^{-1}$ all the data in the range $\tilde D \le 10$ collapse on the same plot.
    }
    \label{fig:63}
\end{figure}

The real and imaginary part of $m_s$ being related as (see eq \ref{eq_48}):
$$\Im m_s=-\frac{\omega_c}{\omega b(\tilde z)}\frac{d}{d\tilde z} \left ( (\tilde z-\tilde D)h(\tilde z) \frac{d \Re m_s}{d\tilde z}\right )
$$
the diffusio-kinetic damping converges at high frequency towards a function of $\tilde D$ strictly proportional to $\omega_c/\omega$. This property is evidenced in figure \ref{fig:63} left, showing that the quantity $(\omega/\omega_c)\Im{\mathcal Z_{DK}} $ reaches a frequency-independant limit at high frequencies. The dependancy in $1/\omega$ of $\Im{\mathcal Z_{DK}} $ stems from the fact that at high frequency the transport of ions is "frozen" and reversible as it follows the surfaces motion, thus the damping ultimately vanishes. 
The convergence of $(\omega/\omega_c)\Im{\mathcal Z_{DK}} $ towards its high-frequency limit follows fully the scheme of the diffusio-kinetic stiffness:  the limit is essentially obtained for $\omega/ \kappa \omega_c \ge 1$,   and the departure from the limit occurs first at small distances  $D < D_s/R\omega$, as was observed on the stiffness (see figure \ref{fig:63} left). 

In the lower frequency range,  the damping reverses its frequency variation and becomes proportional to  $\omega$. It thus recovers the usual damping feature of being proportional to the forced velocity. As shown on figure \ref{fig:63} right, the quantity $(\omega_c/\omega)\Im{\mathcal Z_{DK}} $ which governs the relative damping factor $ \tilde f_{DK}$ (see equation \ref{eq_412}) reaches a low frequency limit $ [(\omega_c/\omega)\Im{\mathcal Z_{DK}}]^{lf}$ when $\omega \rightarrow 0$. This low frequency limit appears to be an upper limit  over all the distance range, and for all parameters $(\alpha,\kappa)$. Therefore the highest
possible values of the diffusion-kinetic damping factor $ \tilde f_{DK}$ are always obtained when the convergence of $(\omega_c/\omega)\Im{\mathcal Z_{DK}} $ to its low-
frequency limit is reached.

Figure \ref{fig:63} right shows the typical features of the convergence. When $\omega$ is decreased, the small distance range converges first, before the high distance range. For most parameters 
the full convergence is obtained for $\omega/\omega_c \le 0.1 \kappa$ in the distance range $\tilde D \le 10$. Thus the diffusio-kinetic damping factor $\tilde f_{DK}$ reaches its maximum value $\tilde f_{DK}^{lf}$ at $\tilde D \le 10$ for $\omega/\omega_c \le 0.1 \kappa$.
\begin{figure}              
       \includegraphics[width=0.7\textwidth]{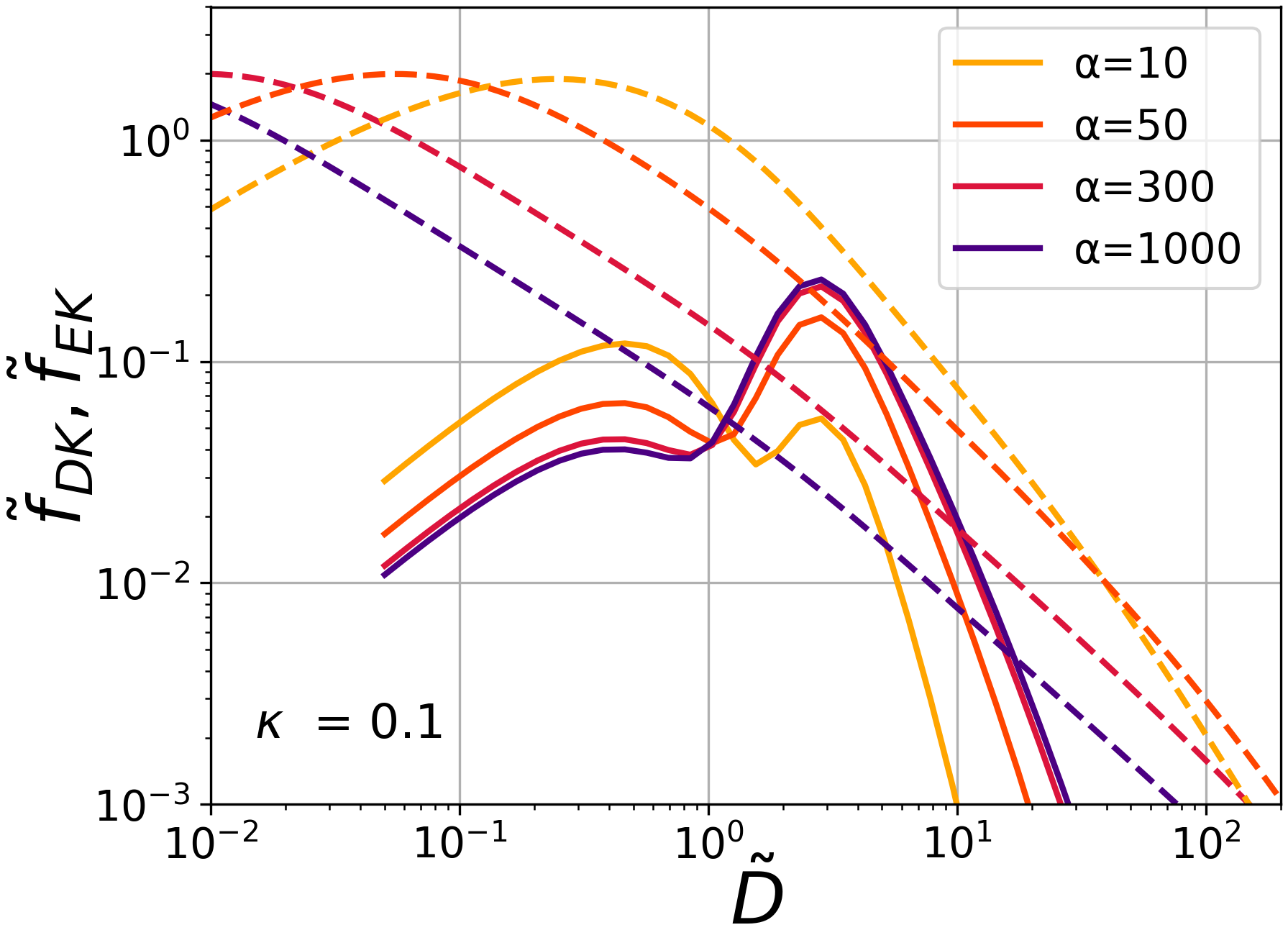}
       \includegraphics[width=0.7\textwidth]{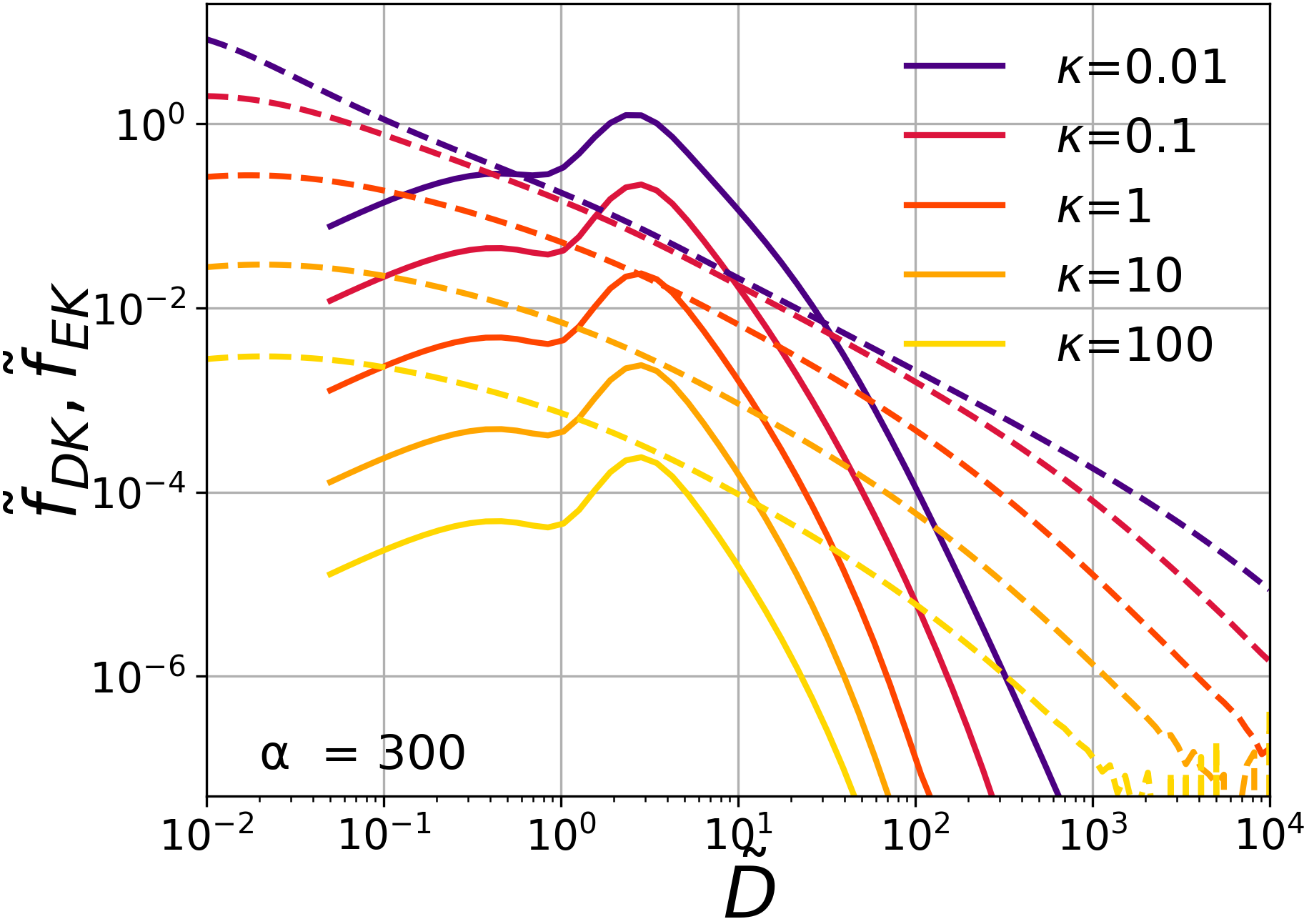}
    \caption{Left: imaginary part of the diffusio-kinetic impedance multiplied by the frequency ratio $\Im(\mathcal Z_{DK})*\omega/\omega_c$  as a function of the adimensional gap $\tilde D=D/2\ell_D$, for $\alpha=10$, $\kappa=0.1$ and various values of $\omega/\kappa \omega_c$. For values $\omega/\kappa \omega_c\ge 10$ all the data collapse on the black dashed line. Right: imaginary part of the diffusio-kinetic impedance  $\Im(\mathcal Z_{DK})$ as a function of $\tilde D$ for the same values of $\alpha$, $\kappa$ and $\omega/\kappa \omega_c$ as on the left graph. 
    }
    \label{fig:64}
\end{figure}

\smallskip
{\em \parindent=0pt Comparison to the Reynolds damping}
\smallskip

In the following we compare the diffusio-kinetic damping to the Reynolds damping by plotting the relative factor $\tilde f_{DK}=(\omega_c/\omega)\Im{\mathcal Z_{DK}}\tilde D/6$ at its maximum value, that is at frequencies $\omega = 0.1 \omega_c\kappa$ or lower.

Figure \ref{fig:64} top plots the diffusio-kinetic factor $\tilde f_{DK}$
obtained for various surface charges at a typical $\kappa=0.1$. The factor shows two peak values at distances $\tilde D \sim 0.4$ and $\tilde D \sim 3$, the second being more important in amplitude. The amplitude of the second peak increases with the surface charge, and saturates for large values $\alpha \ge 100$, where the maximum diffusio-kinetic damping reaches about 30\% of the Reynolds damping. For comparison with the electro-kinetic damping, the factor $\tilde f_{EK}$ is plotted in dashed-line in figure \ref{fig:64}. The diffusio-kinetic damping is in general significantly lower than the electro-kinetic one. However it is of interest to note that due to the charge inversion effect in the electro-kinetic damping, the diffusio-kinetic factor $\tilde f_{DK}$ may become comparable  and even dominant  over $\tilde f_{EK}$,  for $\alpha \ge 50$ and in the distance range $1 \le \tilde D \le 10$. 

Figure \ref{fig:64} down plots the diffusio-kinetic factor $\tilde f_{DK}$
at the rather high value of $\alpha=300$  for various values of  $\kappa$. We observe that $\tilde f_{DK}$ increases as $\kappa$ decreases, a trend already observed in the electro-kinetic damping and due to the restauring character of molecular diffusion. However the variation of $\tilde f_{DK}$ is very regular in $1/\kappa$,  and does not show an intermediate distance range of large Dushin number where the effect of diffusion saturates at low $\kappa$, as it occurs for the electro-kinetic damping. Therefore, the diffusio-kinetic factor exceeds the electro-kinetic factor at $\kappa \ge 1$ in the range $2\ell_D < D < 20 \ell_D$, and may reach 100\% of the Reynolds damping at $\kappa = 0.01$.

In summary the main features of the diffusio-kinetic impedance are: \\
$\bullet$  a long-range stiffness decaying in $1/D^4$, frequency-independant for $\omega \ge D_s/R\ell_D$, and susceptible to overcome the DLVO stiffness at $\tilde D \gtrsim   4$  \\
$\bullet $ a damping in most of the cases negligible, but that can overcome the electro-kinetic damping and even reach the level of the Reynolds damping if three conditions are met: a low frequency $\omega \le 0.1 D_s/R\ell_D$, a high parameter $\alpha=2\ell_D/\ell_G \le 50$ (i.e. nominal surface potential $\psi_s \sim$ 200 mV), and ions of large hydrodynamic radius $r_H \le \ell_B$.


\section{Conclusion}

In this paper, we develop a semi-analytical model for hydrodynamic forces in electric double layers within the linear response regime, considering the drainage of a thin liquid film confined by dielectric solids. We assume the film to be at equilibrium along its thickness, which holds for an ionic aqueous solution at a concentration of $\SI{10}{\micro \Molar}$ or greater ($\lambda_\mathrm{D} = \SI{100}{\nano \meter}$ or lower) when the working frequency is small compared to $\SI{100}{\kilo \hertz}$. \\
Unlike Mugele and collab., who do not take into account the transport of ions by convection in their model – which can be legitimate when using microscopic probes –  we consider here all coupled transports of volume, charge and solute \citep{Mugele2020}. Hence, we find that, when the lateral diffusion of ions cannot follow the excitation frequency, convection takes over and the hydrodynamic force exhibits a conservative component arising from the emerging diffusio-osmotic flow. Yet, this conservative diffusio-osmotic force, which had never been calculated before, may be significant in experiments compared to the DLVO force when ionic diffusion is low, as in ionic liquids. \\
Futhermore, we provide an analytical calculation of the additional electro-hydrodynamic viscous force with respect to a dielectric liquid, of which Mugele and collab. has provided a semi-analytical calculation from an integral expression. We also find that the additional viscous force stemming from diffusio-osmotic effects is in most cases negligible with respect to the previous one. \\
We find that the additional electro-hydrodynamic viscous force does not vary monotonically with the surface charge and vanishes for highly charged surfaces. We demonstrate that this highly counterintuitive effect arises from the divergence of the conductivity of electrolyte liquid films at small thicknesses due to the contribution of counterions. More intuitively, this additional force increases as ion diffusion decreases, with ions playing an asymmetric role depending on the sign of their charge: the diffusion of counter-ions – relative to the sign of the surface charge – predominates that of co-ions. \\
Finally, our theory provides a means to measure the surface charge governing diffusio-electrokinetic transport phenomena – assumed to be constant and unique, though these constraints can be relaxed – which can then be directly compared to the DLVO surface charge without any adjustable parameters. We therefore believe that a direct confrontation of this model with experiments could lead to a better understanding of what the ill-defined concept of surface charge encompasses.

\medskip
{\bf Acknowledgements}.  We thank Lydéric Bocquet for many helpfull discussions. We thank Abdelhamid Maali for pointing to us the work of \cite{Prieve1990}.

\medskip
{\bf Funding}. This work has been funded by the french Agence Nationale pour la Recherche under the grant number ANR-19-CE30-0012-01.


\appendix
\section{Consequences of the Derjaguin approximation}\label{A}
\subsection{Electric field in the dielectric solids}
The potential $\psi$ in the dielectric spheres satisfies the harmonic equation $\Delta \psi=0$ with the boundary condition $\psi=\psi_s$ at the surface of the spheres. Far from the polar axis, $\psi_s$ is uniform, determined by the surface potential of the solid immersed in the electrolyte solution, and  $\psi$ is also uniform. Close to the $Oy$ axis the two spheres may come into electrostatic interaction, and $\psi_s(r)$ is no longer constant. But in the Derjaguin approximation this occurs only in a region where the sphere surfaces are almost parallel, that is perpendicular to $Oy$. Therefore the harmonic equation is solved in the half-space $y\ge z/2$, neglecting the variation of $z$ with $r$. The solution  is:
$$\psi(r,y)= \int_0^\infty e^{-s(y-z/2)} \tilde \psi_s(s)J_0(sr) s \ ds \qquad \tilde \psi_s(s)=\int_0^\infty J_0(sr) \psi_s(r) r dr
$$
where $J_0$ is the Bessel function of the first kind and 0th order. The normal and tangential fields in the sphere at the level of the surface are: 
$$ \frac{\partial \psi}{\partial y}\vert_{z/2^+}= -\int_0^\infty  \tilde \psi_s(s)J_0(sr) s^2 \ ds
\qquad \frac{\partial \psi}{\partial r}\vert_{z/2} = -\int_0^\infty  \tilde \psi_s(s)J_1(sr) s^2 \ ds$$
Therefore the normal electric field inside the sphere is of same  magnitude as the tangential electric field. The latter is negligible in the electrolyte at the level of the surface, due to the Derjaguin approximation. Thus in equation (\ref{eq_bcE}) the normal field inside the sphere can be neglected. 
\subsection{General expression of the force}
We start from equation (\ref{eq_215}), and note that  due to the flow incompressibility and the no-slip boundary condition, at $y=z/2$ we have $ \partial v_y/\partial y=- (1/r)( \partial (rv_r)/\partial r)=0$.

Taking the $y$ component of the Stokes equation (\ref{eq_Stokes}), replacing in it $(n^+-n^-)$ according to the Poisson equation (\ref{eq_Poissonaxial}), and neglecting the radial component $\Delta_r v_y$, we obtain the relation: 
\begin{equation}
0=-\frac{\partial P}{\partial y}+\eta \frac{\partial ^2 v_y}{\partial y^2}+\frac{k_BT}{4\pi \ell_B}\frac{\partial ^2 \psi}{\partial y^2}\frac{\partial  \psi}{\partial y}
\label{eq_216}
\end{equation}
(\ref{eq_216}) is integrated along $y$  with the boundary conditions (\ref{eq_bcsym}) and the incompressibility relation to find:  
\begin{align}
\left [-P+ \frac{k_BT}{8\pi \ell_B}\left ( \frac{\partial \psi}{\partial y} \right )^2 \right ] \left (r, \left (\frac{z}{2} \right)^-\right )  = - \left [P+ \frac{\eta}{r} \frac{\partial  rv_r}{\partial r} \right ] (r, 0 )
\end{align}
so that  the force (\ref{eq_215}) resumes to:
\begin{equation}
    F=\int_0^\infty 2\pi r P(r,0)dr + 2\pi \eta \lim_{r\rightarrow \infty} [rv_r(r,0)]
\end{equation}
The magnitude of the second term in the above expression can be estimated from
$$\displaystyle \eta\vert rv_r \vert \propto \eta\vert r \frac{dv_r}{dr}\vert = \eta\vert r^2 \frac{dv_y}{dy}\vert \propto \eta\vert \frac{r^2\dot D}{z} \vert \rightarrow \eta R\dot D \quad {\rm when} \ r \rightarrow \infty$$
This term is negligible with respect to the first one whose magnitude is of order $\displaystyle \eta \frac{R^2 \dot D}{D}$. 

\subsection{Invariance of $\Pi$ across the film thickness}
By injecting equation (\ref{eq_38}) in the Stokes equation (\ref{eq_Stokes}) we get
$$\eta \nabla^2 \vec v =\vec \nabla (P-k_BTn_s) +en_e \vec \nabla W+n_s \vec \nabla \mu_s \simeq \eta \frac{\partial^2 v_y}{\partial y^2}
$$
as $\Delta_r \ll \partial^2/\partial y^2$. The fields $W$ and $\mu_s$ and their gradient are uniform in the section, therefore projecting the above equation on $y$ reads $\partial \Pi / \partial y = \eta \partial^2 v_y / \partial y^2   $. By integrating this equation over $y$  and derivating the result against $r$, taking into account $\partial v_y/\partial y = - (1/r)(\partial v_r/\partial r)$, we see that the magnitude of the variation of $\partial \Pi /\partial r$ in the $y$ direction is of the order of $\eta \Delta_r v_r$. Therefore at the level of the current approximations   the field $\partial \Pi /\partial r$ is uniform across the liquid film in eq. (\ref{eq_37d}).

\section{Elements of the transport matrix}\label{B}
From equations (\ref{eq_37b},\ref{eq_37c}) one defines the diffusive part of the concentration and charge fluxes as:
$$J_{\rm s,diff}=-\frac{D_s}{k_BT}\frac{d \mu_{\rm s}}{dr}\int_0^{z(t)/2}(n_{\rm s} +n_{\rm e}\delta )dy -\frac{eD_s}{k_BT}\frac{d W}{dr}\int_0^{z(t)/2}(n_{\rm e} +n_{\rm s}\delta )dy
$$
$$J_{\rm s,diff}=-\frac{D_s}{k_BT}\frac{d \mu_{\rm s}}{dr}[N_s+n_oz+N_e\delta] -\frac{eD_s}{k_BT}\frac{d W}{dr}(N_e+(N_s+n_oz)\delta) $$
$$J_{\rm e,diff}=-\frac{D_s}{k_BT}\frac{d \mu_{\rm s}}{dr}(N_e+(N_s+n_oz)\delta) -\frac{eD_s}{k_BT}\frac{d W}{dr}((N_s+n_oz+N_e\delta) $$
One then introduces $v'_r=(\eta/2n_ok_BT)v_r$ and integrate (\ref{eq_37d})  over $y$ using the symmetry boundary condition (\ref{eq_bcsym}) $\partial v_r(y)/\partial y (y=0)=0$:
$$\eta \frac{\partial v_r(r,y,t)}{\partial y}=\Gamma_{\rm v} \frac{d \Pi}{d r}+ \Gamma_e(r,y,t)\frac{d W}{d r} + \Gamma_s(r,y,t)\frac{d \mu_s}{d r}
$$
$$\Gamma_{\rm v}=\int_0^y dy=y \qquad \Gamma_s(r,y,t)=\int_0^{y}(n_s-2c)dy  \qquad \Gamma_e=\int_0^{y}en_e dy
$$
The volume flux can be expressed as:
$$J_{\rm v}=\int_0^{z(t)/2}v_r(y)dy=-\int_0^{z(t)/2}y\frac{\partial v_r(y)}{\partial y}dy
$$
$$J_{\rm v}=-\frac{1}{\eta}\left [ \frac{d \Pi}{d r}\int_0^{z(t)/2}\Gamma^2_{\rm v}dy   +\frac{d \mu_s}{d r}\int_0^{z(t)/2}\Gamma_{\rm v} \Gamma_s(y)dy +  \frac{d W}{d r} \int_0^{z(t)/2}\Gamma_{\rm v} \Gamma_e(y)dy
\right]
$$
The convective charge flux writes:
$$I=eJ_{\rm e}=\int_0^{z(t)/2}en_ev_r(y)dy=-\int_0^{z(t)/2}\Gamma_e\frac{dv_r}{dy}dy
$$
and the convective ion flux:
$$J_{\rm s,conv}=2c J_{\rm v}+\int_0^{z(t)/2}(n_s-2c)v_r(y)=2c J_{\rm v}-\int_0^{z(t)/2}\Gamma_{\rm s}\frac{dv_r}{dy}dy
$$
Altogether the fluxes $(J_{\rm v}, J_{\rm s}-cJ_{\rm v}, I)$ are related to the gradients of the potentials $\Pi, \mu_{\rm s}$ and $W$ by a symmetric transport matrix $\mathsfbi{T} $ according to: 
$$\begin{pmatrix}J_{\rm v} \\J_{\rm s}-cJ_{\rm v} \\ I \end{pmatrix} =-\mathsfbi{T} \frac{d}{dr}
\begin{pmatrix}\Pi \\ \mu_s \\  W\end{pmatrix} \qquad 
T_{ij}=\int_0^{z(t)/2}\frac{\Gamma_i\Gamma_j}{\eta}dy+\frac{D_s}{k_BT}L_{ij}
$$
$$L_{\rm v i}=0 \qquad L_{\rm ss}=\frac{L_{\rm ee}}{e^2}=N_s+n_oz+N_e\delta \quad L_{\rm es}=L_{\rm se}=e(N_e+(N_s+n_oz)\delta)
$$

Introducing the local time-dependent Debye's length $\lambda_D=\ell_De^{-\mu_s/2k_BT}$, and the non-dimensional lengths $\tilde y=y/2\lambda_D$, the functions $\Gamma_{s,e}$ express as
$$\Gamma_{s,e}=4c\lambda_D \tilde \Gamma_{s,e} \qquad \tilde \Gamma_s=\int_0^{\tilde y}(\cosh \chi-1) d\tilde y' \quad \tilde \Gamma_e=-\int_0^{\tilde y}\sinh \chi d\tilde y' 
$$
where $\chi$ is the function $\psi_{eq}$   calculated with the boundaries located at $\pm z(t)/2$ and with the local Debye's length $\lambda_D$.
From (\cite{Abramowitz1964}) the  $\tilde \Gamma_{se}$ express as
\begin{IEEEeqnarray}{rCl}
\Gamma_{s}(\tilde y)=\frac{1}{2\sqrt{k}}\left (  2u(1-k)-2E(u,k^2)+{\rm sn} (u,k^2)(k^2{\rm cd}(u,k^2)+\frac{1}{{\rm cd}(u,k^2)}\right )
\quad 
\IEEEyesnumber \label{Gamma_exp} \IEEEyessubnumber\label{Gammas_exp} 
\\
\Gamma_e(\tilde y)=-\frac{1-k^2}{2\sqrt{k}}\frac{{\rm sn} (u,k^2)}{{\rm cn} (u,k^2){\rm dn} (u,k^2)} \quad u=\frac{y}{2\lambda_D \sqrt{k}} \qquad k=e^{-\chi_m} \qquad
\IEEEyessubnumber \label{Gammae_exp} \\
\frac{\alpha \sqrt{k}}{1-k^2}e^{-\mu_s/2k_BT}=\frac{{\rm sn} (\tilde z(t)/2\sqrt{k},k^2)}{{\rm cn} (\tilde z(t)/2\sqrt{k},k^2){\rm dn} (\tilde z(t)/2\sqrt{k},k^2)} \quad
\IEEEyessubnumber \label{cl}
\end{IEEEeqnarray}
where $E(u,k^2)=E(\phi \vert \ k^2)$ is the  incomplete elliptic integral of the second kind of parameter $k^2$ associated to the  elliptic integral of the first kind $F(\phi \vert \ k^2)=u$, the functions sn, cn, dn are Jacobi elliptic functions, and \ref{cl} is the boundary condition at wall.

\section{Expression of $dN_s$ in the linear response}\label{C}
The ion excess  $ N_s$ is
$$ N_s=2c\int_0^{z(t)/2}(\cosh \chi -1)dy +(2c-2n_o)\frac{z(t)}{2}$$
\begin{equation}
    \frac{N_s}{4n_o \ell_D}= e^{\mu_s/2k_BT}\tilde \Gamma_s(\frac{\tilde z(t)}{4\lambda_D\sqrt{k}})+\frac{\tilde z(t)}{2}(e^{\mu_s/k_BT}-1)
    \label{eq_Ns}
\end{equation}
The variation of $N_s$ with $z(t)$ and $\mu_s$ around equilibrium ($\mu_s=0$) has to take into account the variation of $k$ with these quantities. The latter is obtained by differentiating the boundary condition \ref{cl}. Introducing the equilibrium potential at the wall $\psi_{\rm eq,s}$ defined by the boundary condition (\ref{cl}) at equilibrium:
$$\alpha=\frac{2\ell_D}{\ell_G} \qquad \cosh \psi_{\rm eq,s}=\cosh \psi_{\rm eq,m}+\frac{\alpha^2}{2}
$$
one obtains
\begin{equation}\frac{dk}{k}=-\frac{\frac{h_0}{\ell_D}+\frac{\mu_s}{k_BT}(\tilde z+\alpha/2\sinh \psi_{\rm s})}{A } \qquad A=\tilde z-\frac{2E(\tilde z/2\sqrt{k})}{\sqrt{k}\sinh \psi_m}- \frac{2}{\alpha}(1-\frac{\sinh \psi_{\rm m}}{\sinh \psi_{\rm s}})
\end{equation}
Then by differentiating (\ref{eq_Ns}) and using the reduced field 
$\displaystyle m_s=\frac{\mu_s \ell_D}{k_BT h_0}$  one obtains:
\begin{IEEEeqnarray}{rCl}
\frac{dN_s}{2n_oh_0} & = & \frac{a(\tilde z)}{2}+ m_s\frac{b(\tilde z)}{2}
   \IEEEyesnumber  \IEEEyessubnumber\\
    a(\tilde z) & = & \cosh \psi_{\rm m}-1+\frac{\sinh \psi_{\rm m}}{A(\tilde z)}(\tilde z+\frac{\alpha}{\sinh \psi_{\rm s}}) \IEEEyessubnumber\\
    b(\tilde z) & = &\frac{\tilde z}{2}(\frac{3}{k}+k)-\frac{2}{\sqrt{k}}E(\frac{\tilde z}{2\sqrt{k}})-\frac{\alpha^3}{2\sinh \psi_{\rm s}}+\frac{2}{\alpha}(\sinh \psi_{\rm s}-\sinh \psi_{\rm m}) \nonumber \\
    && +\: \frac{\sinh \psi_{\rm m}}{A(\tilde z)}(\tilde z+\frac{\alpha}{\sinh \psi_{\rm s}})^2
    \IEEEyessubnumber
\end{IEEEeqnarray}
Here the potentials are equilibrium potentials and $\tilde z=z/2\ell_D$.

Finally when the EDL's do not overlap, $\sinh \psi_m=0$, $\cosh \psi_m=1=k=E(z/4\ell_D\sqrt{k})$, $\sinh \psi_s=\alpha\sqrt{1+\alpha^2/4}$. Introducing $\gamma=\tanh \psi_s/4$  one finds that $dN_s$ reduces to: 
\begin{equation}
    \frac{dN_s}{2n_o h_0}=m_s \left (\tilde z+\frac{1}{\sqrt{1+\ell_D^2/\ell_G^2}}-1 \right )=m_s \left (\tilde z-\frac{2\gamma^2}{1+\gamma^2} \right )
\end{equation}
\begin{equation}
 a(\tilde z)=0 \qquad   b(\tilde z)=2(\tilde z-\frac{2\gamma^2}{1+\gamma^2})
\end{equation}


\section{Transport coefficients in the non-overlapping regime}\label{D}
Non-overlapping EDL's are treated as single surfaces in solutions, with the Gouy-Chapmann theory giving the reduced potential $\psi_{eq}$ solution of the Poisson-Boltzmann equation with vanishing potential in the mid-plane $\psi_{eq,m}=0$. In the following we write $\psi_{eq}=\psi$ to lighten the expressions and note $\tilde y=y/2\ell_D$:
\begin{IEEEeqnarray}{rCl}
\dfrac{\mathrm{d}\psi}{\mathrm{d} \tilde y}  = 4 \sinh \frac{\psi}{2} \qquad  \sinh \frac{\psi_s}{2}=\frac{\ell_D}{\ell_G}=\frac{\alpha}{2}=\frac{2\gamma}{1-\gamma^2} \\
t=\tanh \frac{\psi(\tilde y)}{4} = \gamma \exp (2\tilde y - \tilde z) \qquad
\gamma = \tanh \frac{\psi_s}{4} = \frac{2}{\alpha} \left(\sqrt{1+ \frac{\alpha^2}{4}} -1 \right)  \\
\tilde \Gamma_s(\tilde y) = \int_0^{\tilde y}(\cosh \psi-1) {\rm d}\tilde y' = \cosh \frac{\psi}{2}-1 \qquad  
\tilde \Gamma_e(\tilde y) =-\sinh \frac{\psi}{2}=-\dfrac{d\psi}{4d \tilde y}
\end{IEEEeqnarray}
With these expressions it is possible to perform analytically the integrals $\int_0^{\tilde z/2} \tilde y\tilde \Gamma_{s,e} d\tilde y  $ and $\int_0^{\tilde z/2} \tilde \Gamma_{s,e}\tilde \Gamma_{s,e} d\tilde y  $ and we find $t_{ij}=a_{ij}\tilde z+b_{ij} $ (except for $t_{\rm vv}=\tilde z/12$) as:
\begin{IEEEeqnarray}{rCl}
a_{{\rm ss}}=  a_{{\rm ee}} =\kappa \qquad a_{{\rm se}}=\kappa\delta \qquad a_{{\rm vs}}=-\frac{1}{2}\ln (1-\gamma^2) \qquad \\
a_{{\rm ve}}=-\frac{1}{2}\ln \frac{1+\gamma}{1-\gamma}=-\frac{\psi_s}{4} \quad 
\IEEEyesnumber \IEEEyessubnumber \\
b_{{\rm ve}}=\frac{1}{2}({\rm Li}_2(\gamma)-{\rm Li}_2(-\gamma) \qquad b_{{\rm vs}}=-\frac{{\rm Li}_2(\gamma^2)}{4} \quad 
\IEEEyessubnumber \\
 b_{{\rm ee}}=\frac{2\gamma^2}{1-\gamma^2}(1+2\kappa(1-\frac{\delta}{\gamma}))
 \qquad 
b_{{\rm se}}=\ln\frac{1+\gamma}{1-\gamma}-\frac{2\gamma}{1-\gamma^2}[1+2\kappa(1-\delta \gamma)] \quad 
\IEEEyessubnumber \\
 b_{{\rm ss}}=2\ln (1-\gamma^2)+\frac{2\gamma^2}{1-\gamma^2} (1+2\kappa(1-\frac{\delta}{\gamma}))\quad
\IEEEyessubnumber 
\end{IEEEeqnarray}
where $\displaystyle {\rm Li}_2(x) = -\int_0^x \frac{\ln(1-u)}{u}{\rm d}u$ and $\displaystyle \mathrm{Li}_2(1-x)=f(x)=-\int_1^x \frac{\ln u}{u-1}{\rm d}u$ 
 is the dilogarithm function $f$ defined by \cite{Abramowitz1964} page 1005 paragraph 27.7.

\section{Donnan's model}\label{E}
A simplified model for the thin electrolyte film is the so-called Donnan's model \cite{Donnan1924} whereby the ions concentration and the electrical potential are considered uniform in the thickness, satisfying globally the Boltzmann law $n^{\pm }_{eq} =n_oe^{\pm \psi_D} $ and the electroneutrality condition. The Donan potential $\psi_{D,eq}(z)$ in the electrolyte film of thickness $z$ is given by 
\begin {equation}
\sinh \psi_{D,eq}(z) = {\rm Du} = \frac{\sigma}{en_oz}=\frac{\alpha}{\tilde z}
\label{eq_E1}
\end{equation}
Within this model the quantities $\tilde \Gamma_{s,e}$ resume to $\tilde \Gamma_s(\tilde y)=\tilde y (\cosh \psi_{D,eq}(z)-1)$, $\tilde \Gamma_e(\tilde y)=-\tilde y \sinh \psi_{D,eq}(z)$, and the transport coefficient $t_{ij}$ write as: 
\begin{eqnarray}
t_{\rm ve}=-\sinh \psi_{D}\frac{\tilde z^3}{12}=-\frac{\alpha \tilde z^2}{12} \qquad t_{\rm vs}=(\cosh \psi_{D}-1)\frac{\tilde z^3}{12} \\
t_{\rm ee}=\sinh^2\psi_{D}\frac{\tilde z^3}{12}+\kappa \tilde z(\cosh \psi_D-\delta \sinh \psi_D)=\frac{\alpha^2 \tilde z}{12}+\kappa(\tilde z \cosh \psi_D-\delta \alpha) \\ 
t_{\rm ss}=(\cosh \psi_D-1)^2\frac{\tilde z^3}{12}+\kappa (\tilde z\cosh \psi_D-\delta \alpha) \\
t_{\rm se}=-\cosh \psi_D\frac{\alpha \tilde z^2}{12}+\kappa (\tilde z\delta \cosh \psi_D- \alpha)
    \end{eqnarray}
The Donan's model is considered to reflect adequately the transport properties when the Dushin number is larger than 1, that is when $\tilde z$ is sufficiently small so that $\alpha/\tilde z \ge 1$. The transport coefficients can be further simplified  as:
\begin{equation}
    t_{\rm vs}\simeq -t_{\rm ve}=\frac{\alpha z^2}{12} \quad t_{\rm ee}\simeq t_{\rm ss}\simeq -t_{\rm es}=\frac{\alpha^2 z}{12}+\kappa \alpha(1-\delta)
\end{equation}


\bibliographystyle{jfm}
\bibliography{jfm-instructions}

\begin{thebibliography}{31}
\expandafter\ifx\csname natexlab\endcsname\relax\def\natexlab#1{#1}\fi
\def\au#1{#1} \def\ed#1{#1} \def\yr#1{#1}\def\at#1{#1}\def\jt#1{\textit{#1}} \def\bt#1{#1}\def\bvol#1{\textbf{#1}} \def\vol#1{#1} \def\pg#1{#1} \def\publ#1{#1}\def\arxiv#1{#1}\def\org#1{#1}\def\st#1{\textit{#1}}

\bibitem[Abramowitz \& Stegun(1964)]{Abramowitz1964}
{\sc \au{Abramowitz, M.} \& \au{Stegun, I.A.}} \yr{1964} {\em Handbook of Mathematical Functions with Formulas, Graphs, and Mathematical Tables\/}.  \publ{U.S. Government Printing Office}.

\bibitem[Anderson(1989)]{Anderson1989}
{\sc \au{Anderson, J.L.}} \yr{1989}  \at{Colloid transport by interfacial forces}.  \jt{Annu. Rev. Fluid Mech.}  \bvol{21},  \pg{61--99}.

\bibitem[Behrens \& Grier(2001)]{Behrens2001}
{\sc \au{Behrens, S.H.} \& \au{Grier, D.G.}} \yr{2001}  \at{Giant osmotic energy conversion measured in a single transmembrane boron nitride nanotube}.  \jt{J Chem Phys}  \bvol{115},  \pg{6716}.

\bibitem[Bike \& Prieve(1990)]{Prieve1990}
{\sc \au{Bike, S.~G.} \& \au{Prieve, D.~C.}} \yr{1990}  \at{Electrohydrodynamic lubrication with thin double layers}.  \jt{Journal of Colloid and Interface Science}  \bvol{136 (1)},  \pg{95–112}.

\bibitem[Bocquet(2020)]{Bocquet2020}
{\sc \au{Bocquet, L.}} \yr{2020}  \at{Nanofluidics coming of age}.  \jt{Nat. Mater.}  \bvol{19 (3)},  \pg{254–256}.

\bibitem[Bocquet \& Barrat(2007)]{Bocquet2007}
{\sc \au{Bocquet, L.} \& \au{Barrat, J.-L.}} \yr{2007}  \at{Flow boundary conditions from nano- to micro-scales}.  \jt{Soft Matter}  \bvol{3},  \pg{685–693}.

\bibitem[Bocquet \& Charlaix(2010)]{Bocquet2010}
{\sc \au{Bocquet, L.} \& \au{Charlaix, E.}} \yr{2010}  \at{Nanofluidics, from bulk to interfaces}.  \jt{Chemical Society Reviews}  \bvol{39 (3)},  \pg{1073–1095}.

\bibitem[Bonthuis \& Netz(2012)]{Bonthuis2012}
{\sc \au{Bonthuis, D.~J.} \& \au{Netz, R.~R.}} \yr{2012}  \at{Unraveling the combined effects of dielectric and viscosity profiles on surface capacitance, electro-osmotic mobility, and electric surface conductivity}.  \jt{Langmuir}  \bvol{28 (46)},  \pg{16049–16059}.

\bibitem[Bourg {\em et~al.\/}(2017)Bourg, Lee, Fenter \& Tournassat]{Bourg2017}
{\sc \au{Bourg, I.~C.}, \au{Lee, S.~S.}, \au{Fenter, P.} \& \au{Tournassat, C.}} \yr{2017}  \at{Stern layer structure and energetics at mica–water interfaces}.  \jt{J. Phys. Chem. C}  \bvol{121 (17)},  \pg{9402–9412}.

\bibitem[Cottin-Bizonne {\em et~al.\/}(2005)Cottin-Bizonne, Cross, Steinberger \& Charlaix]{Cottin-B2005}
{\sc \au{Cottin-Bizonne, C.}, \au{Cross, B.}, \au{Steinberger, A.} \& \au{Charlaix, E.}} \yr{2005}  \at{Boundary slip on smooth hydrophobic surfaces: Intrinsic effects and possible artifacts}.  \jt{Phys. Rev. Lett.}  \bvol{94 (5)},  \pg{056102}.

\bibitem[Donnan(1924)]{Donnan1924}
{\sc \au{Donnan, F.G.}} \yr{1924}  \at{The theory of membrane equilibria}.  \jt{Chem. Rev.}  \bvol{1},  \pg{73--90}.

\bibitem[Garcia {\em et~al.\/}(2016)Garcia, Barraud, Picard, Giraud, Charlaix \& Cross]{Garcia2016}
{\sc \au{Garcia, L.}, \au{Barraud, C.}, \au{Picard, C.}, \au{Giraud, J.}, \au{Charlaix, E.} \& \au{Cross, B.}} \yr{2016}  \at{Micro-nano-rheometer for the mechanics of soft matter at interfaces}.  \jt{Review of Scientific Instruments}  \bvol{87 (11)},  \pg{113906}.

\bibitem[Gonella {\em et~al.\/}(2021)Gonella, Backus, Nagata, Bonthuis, Loche, Schlaich, Netz, Kühnle, McCrum, Koper, Wolf, Winter, Meijer, Campen \& Bonn]{Bonn2021}
{\sc \au{Gonella, G.}, \au{Backus, E. H.~G.}, \au{Nagata, Y.}, \au{Bonthuis, D.~J.}, \au{Loche, P.}, \au{Schlaich, A.}, \au{Netz, R.~R.}, \au{Kühnle, A.}, \au{McCrum, I.~T.}, \au{Koper, M. T.~M.}, \au{Wolf, M.}, \au{Winter, B.}, \au{Meijer, G.}, \au{Campen, R.~K.} \& \au{Bonn, M.}} \yr{2021}  \at{Water at charged interfaces}.  \jt{Nat Rev Chem}  \bvol{5 (7)},  \pg{466–485}.

\bibitem[Gross \& Osterle(1968)]{Gross1968}
{\sc \au{Gross, R.~J.} \& \au{Osterle, J.~F.}} \yr{1968}  \at{Membrane transport characteristics of ultrafine capillaries}.  \jt{The Journal of Chemical Physics}  \bvol{49 (1)},  \pg{228–234}.

\bibitem[Hartkamp {\em et~al.\/}(2018)Hartkamp, Biance, Fu, Dufrêche, Bonhomme \& Joly]{Hartkamp2018}
{\sc \au{Hartkamp, R.}, \au{Biance, A.-L.}, \au{Fu, L.}, \au{Dufrêche, J.-F.}, \au{Bonhomme, O.} \& \au{Joly, L.}} \yr{2018}  \at{Measuring surface charge: Why experimental characterization and molecular modeling should be coupled}.  \jt{Current Opinion in Colloid \& Interface Science}  \bvol{37},  \pg{101–114}.

\bibitem[Horn {\em et~al.\/}(1989)Horn, Smith \& Haller]{Horn1989}
{\sc \au{Horn, R.~G.}, \au{Smith, D.~T.} \& \au{Haller, W.}} \yr{1989}  \at{Surface forces and viscosity of water measured between silica sheets}.  \jt{Chemical Physics Letters}  \bvol{162},  \pg{404--408}.

\bibitem[Israelachvili(2011)]{Israelachvili2011}
{\sc \au{Israelachvili, J.~N.}} \yr{2011} {\em Intermolecular and Surface Forces\/}.  \publ{Academic Press}.

\bibitem[Israelachvili \& Adams(1978)]{Israelachvili1978}
{\sc \au{Israelachvili, J.~N.} \& \au{Adams, G.~E.}} \yr{1978}  \at{Measurement of forces between two mica surfaces in aqueous electrolyte solutions in the range 0–100 nm}.  \jt{Journal of the Chemical Society, Faraday Transactions 1: Physical Chemistry in Condensed Phases}  \bvol{74},  \pg{975--1001}.

\bibitem[Liu {\em et~al.\/}(2018{\natexlab{{\em a\/}}})Liu, Klaassen, Zhao, Mugele \& van~den Ende]{Liu2018}
{\sc \au{Liu, F.}, \au{Klaassen, A.}, \au{Zhao, C.}, \au{Mugele, F.} \& \au{van~den Ende, D.}} \yr{2018{\natexlab{{\em a\/}}}}  \at{Electroviscous dissipation in aqueous electrolyte films with overlapping electric double layers}.  \jt{J. Phys. Chem. B}  \bvol{122},  \pg{933–946}.

\bibitem[Liu {\em et~al.\/}(2018{\natexlab{{\em b\/}}})Liu, Kawaguchi, Pierce, Komanicky \& You]{You2018}
{\sc \au{Liu, Y.}, \au{Kawaguchi, T.}, \au{Pierce, M.~S.}, \au{Komanicky, V.} \& \au{You, H.}} \yr{2018{\natexlab{{\em b\/}}}}  \at{Layering and ordering in electrochemical double layers}.  \jt{J. Phys. Chem. Lett.}  \bvol{9 (6)},  \pg{1265–1271}.

\bibitem[Lizée {\em et~al.\/}(2024)Lizée, Coquinot, Mariette, Siria \& Bocquet]{Lizee2024}
{\sc \au{Lizée, M.}, \au{Coquinot, B.}, \au{Mariette, G.}, \au{Siria, A.} \& \au{Bocquet, L.}} \yr{2024}  \at{Anomalous friction of supercooled glycerol on mica}.  \jt{Nat Commun}  \bvol{15 (1)},  \pg{6129}.

\bibitem[Lyklema(1994)]{Lyklema1994}
{\sc \au{Lyklema, J.}} \yr{1994}  \at{On the slip process in electrokinetics}.  \jt{Colloids and Surfaces A: Physicochemical and Engineering Aspects}  \bvol{92 (1-2)},  \pg{41--49}.

\bibitem[Maali {\em et~al.\/}(2008)Maali, Cohen-Bouhacina \& Kellay]{Maali2008}
{\sc \au{Maali, A.}, \au{Cohen-Bouhacina, T.} \& \au{Kellay, H.}} \yr{2008}  \at{Measurement of the slip length of water flow on graphite surface}.  \jt{Applied Physics Letters}  \bvol{92 (5)},  \pg{053101}.

\bibitem[Prieve {\em et~al.\/}(1984)Prieve, Anderson, Ebel \& Lowell]{Prieve1984}
{\sc \au{Prieve, D.~C.}, \au{Anderson, J.~L.}, \au{Ebel, J.~P.} \& \au{Lowell, M.~E.}} \yr{1984}  \at{Motion of a particle generated by chemical gradients. part 2. electrolytes}.  \jt{Journal of Fluid Mechanics}  \bvol{148},  \pg{247–269}.

\bibitem[Rodríguez~Matus {\em et~al.\/}(2022)Rodríguez~Matus, Zhang, Benrahla, Majee, Maali \& Würger]{Rodriguez2022}
{\sc \au{Rodríguez~Matus, M.}, \au{Zhang, Z.}, \au{Benrahla, Z.}, \au{Majee, A.}, \au{Maali, A.} \& \au{Würger, A.}} \yr{2022}  \at{Electroviscous drag on squeezing motion in sphere-plane geometry}.  \jt{Phys. Rev. E}  \bvol{105 (6)},  \pg{064606}.

\bibitem[Siria {\em et~al.\/}(2017)Siria, Bocquet \& Bocquet]{Siria2017}
{\sc \au{Siria, A.}, \au{Bocquet, M.-L.} \& \au{Bocquet, L.}} \yr{2017}  \at{New avenues for the large-scale harvesting of blue energy}.  \jt{Nat Rev Chem 2017}  \bvol{1 (11)},  \pg{1–10}.

\bibitem[Siria {\em et~al.\/}(2013)Siria, Poncharal, Biance, Fulcrand, Blase, Purcell \& Bocquet]{Siria2013}
{\sc \au{Siria, A.}, \au{Poncharal, P.}, \au{Biance, A.-L.}, \au{Fulcrand, R.}, \au{Blase, X.}, \au{Purcell, S.~T.} \& \au{Bocquet, L.}} \yr{2013}  \at{Giant osmotic energy conversion measured in a single transmembrane boron nitride nanotube}.  \jt{Nature}  \bvol{494 (7438)},  \pg{455–458}.

\bibitem[Smoluchowski(1904)]{Smo1924}
{\sc \au{Smoluchowski, M.}} \yr{1904}  \at{Contribution à la théorie de l’endosmose électrique et de quelques phénomènes corrélatifs}.  \jt{J. Phys. Theor. Appl.}  \bvol{3},  \pg{912}.

\bibitem[Tabor \& Winterton(1969)]{Tabor1969}
{\sc \au{Tabor, D.} \& \au{Winterton, R. H.~S.}} \yr{1969}  \at{The direct measurement of normal and retarded van der {W}aals forces.}  \jt{Proceedings of the Royal Society of London. A. Mathematical and Physical Sciences}  \bvol{312},  \pg{435--450}.

\bibitem[Wang {\em et~al.\/}(2024)Wang, Li, Tavakol, Serva, Nener, Parish, Salanne, Warr, Voïtchovsky \& Atkin]{Atkin2024}
{\sc \au{Wang, J.}, \au{Li, H.}, \au{Tavakol, M.}, \au{Serva, A.}, \au{Nener, B.}, \au{Parish, G.}, \au{Salanne, M.}, \au{Warr, G.~G.}, \au{Voïtchovsky, K.} \& \au{Atkin, R.}} \yr{2024}  \at{Ions adsorbed at amorphous solid/solution interfaces form {W}igner crystal-like structures}.  \jt{ACS Nano}  \bvol{18 (1)},  \pg{1181–1194}.

\bibitem[Zhao {\em et~al.\/}(2020)Zhao, Zhang, van~den Ende \& Mugele]{Mugele2020}
{\sc \au{Zhao, C.}, \au{Zhang, W.}, \au{van~den Ende, D.} \& \au{Mugele, F.}} \yr{2020}  \at{Electroviscous effects on the squeezing flow of thin electrolyte solution films}.  \jt{J. Fluid Mech.}  \bvol{888},  \pg{A29}.

\end{thebibliography}

\end{document}